      [1999/12/01 v1.4c Il Nuovo Cimento]
\documentclass[varenna]{cimento}

\usepackage{graphicx}  
\title{{\it Ab initio} no-core shell model calculations for light nuclei}
\author{Petr Navr\'atil}
\institute{Lawrence Livermore National Laboratory, L-414, P.O. Box 808,
             Livermore, CA  94551, USA}
\begin{document}

\maketitle

\begin{abstract}
An overview of the {\it ab initio} no-core shell model is presented. Recent
results for light nuclei obtained with the chiral two-nucleon and three-nucleon
interactions are highlighted. Cross section calculations of capture reactions 
important for astrophysics are discussed. The extension of the {\it ab initio}
no-core shell model to the description of nuclear reactions
by the resonating group method technique is outlined.
\end{abstract}

\section{Introduction}

The major outstanding problem in nuclear physics is to calculate properties of finite nuclei
starting from the basic interactions among nucleons. This problem has two parts. First, the
basic interactions among nucleons are complicated, they are not uniquely defined and there
is evidence that more than just two-nucleon forces are important. Second, the nuclear many-body
problem is very difficult to solve. This is a direct consequence of the complex nature
of the inter-nucleon interactions. Both short-range and medium-range correlations among 
nucleons are important and for some observables long-range correlations also play a 
significant role. Various methods have been used to solve the few-nucleon problem 
in the past. The Faddeev method \cite{Fad60}
has been successfully applied to solve the three-nucleon bound-state
problem for different nucleon-nucleon potentials \cite{CPFG85,FPSS93,NHKG97}.
For the solution of the four-nucleon problem one can employ Yakubovsky's 
generalization
of the Faddeev formalism  \cite{Ya67} as done, e.g., in Refs. \cite{GH93}
or \cite{CC98}.
Alternatively, other methods have also been succesfully used,
such as, 
the correlated hyperspherical harmonics expansion method \cite{VKR95,BLO99}
or the Green's function Monte Carlo method \cite{GFMC}. 
Recently, a benchmark calculation by seven different methods
was performed for a four-nucleon bound state problem \cite{benchmark}.
However, there are few approaches that can be successfully applied 
to systems of more than four nucleons when realistic inter-nucleon interactions are
used.
Apart from the coupled cluster method \cite{CCM,HM99,KoDe04,WlDe05} applicable typically
to closed-shell nuclei, 
the Green's function Monte Carlo method is a prominent approach
capable to solve the nuclear many-body problem with realistic interactions
for systems of up to $A=12$. Another method developed recently applicable
to light nuclei up to $A=16$ and beyond is the {\it ab initio} no-core shell model 
(NCSM) \cite{NaVa00}. In this paper, an overview of this approach is given and 
results obtained very recently are presented. 
Also, future developments, in particular applications 
to nuclear reactions, are outlined.

\section{{\it Ab initio} no-core shell model}

In the {\it ab initio} no-core shell model, we consider a system 
of $A$ point-like non-relativistic nucleons that interact by realistic two- 
or two- plus three-nucleon (NNN) interactions. Under the term realistic two-nucleon (NN)
interactions we mean NN potentials that fit nucleon-nucleon phase shifts 
with high precision up to certain energy, typically up to 350 MeV. A realistic
NNN interaction includes terms related to two-pion exchanges with an intermediate
delta excitation. In the NCSM, all the nucleons are considered active, 
there is no inert core like in
standard shell model calculations. Therefore the ``no-core'' in the name of the approach.
There are two other major features in addition to the employment of realistic 
NN or NN+NNN interactions. The first one is
the use of the harmonic oscillator (HO) basis truncated by a chosen maximal total HO energy
of the $A$-nucleon system. The reason behind the choice of the HO basis is the fact that
this is the only basis that allows to use single-nucleon coordinates and consequently 
the second-quantization representation without violating the translational invariance
of the system. The powerful techniques based on the second quantization and developed 
for standard shell model calculations can then be utilized. Therefore the ``shell model''
in the name of the approach. As a downside, one has to face the consequences of the incorrect
asymptotic behavior of the HO basis. 
The second feature comes as a consequence of the basis truncation.
In order to speed up convergence with the basis enlargement, we construct an effective interaction
from the original realistic NN or NN+NNN potentials by means of a unitary transformation.
The effective interaction depends on the basis truncation and by construction becomes
the original realistic NN or NN+NNN interaction as the size of the basis approaches infinity.
In principle, one can also perform calculations with the unmodified, ``bare'', original 
interactions. Such calculations are then variational with the HO frequency and the basis truncation
parameter as variational parameters.

\subsection{Hamiltonian}

The starting Hamiltonian of the {\it ab initio} NCSM is
\begin{equation}\label{ham}
H_A= 
\frac{1}{A}\sum_{i<j}\frac{(\vec{p}_i-\vec{p}_j)^2}{2m}
+ \sum_{i<j}^A V_{{\rm NN}, ij} + \sum_{i<j<k}^A V_{{\rm NNN}, ijk} \; ,
\end{equation}
where $m$ is the nucleon mass, $V_{{\rm NN}, ij}$ the NN interaction,
$V_{{\rm NNN}, ijk}$ the three-nucleon interaction. In the NCSM, we employ a large
but finite HO basis. Due to properties of the realistic nuclear 
interaction in Eq. (\ref{ham}),
we must derive an effective interaction appropriate for the basis truncation.
To facilitate the derivation of the effective interaction, we modify the
Hamiltonian (\ref{ham}) by adding to it the center-of-mass (CM) HO Hamiltonian
$H_{\rm CM}=T_{\rm CM}+ U_{\rm CM}$, where
$U_{\rm CM}=\frac{1}{2}Am\Omega^2 \vec{R}^2$,
$\vec{R}=\frac{1}{A}\sum_{i=1}^{A}\vec{r}_i$.
The effect of the HO CM Hamiltonian will later be subtracted
out in the final many-body calculation. Due to the translational invariance of the
Hamiltonian (\ref{ham}) the HO CM Hamiltonian has in fact no effect on the intrinsic
properties of the system. 
The modified Hamiltonian can be cast into the form
\begin{eqnarray}\label{hamomega}
H_A^\Omega &=& H_A + H_{\rm CM}=\sum_{i=1}^A h_i + \sum_{i<j}^A V_{ij}^{\Omega,A}
+\sum_{i<j<k}^A V_{{\rm NNN}, ijk} 
= \sum_{i=1}^A \left[ \frac{\vec{p}_i^2}{2m}
+\frac{1}{2}m\Omega^2 \vec{r}^2_i
\right] 
\nonumber \\ 
&&
+ \sum_{i<j}^A \left[ V_{{\rm NN}, ij}
-\frac{m\Omega^2}{2A}
(\vec{r}_i-\vec{r}_j)^2
\right] + \sum_{i<j<k}^A V_{{\rm NNN}, ijk} \; .
\end{eqnarray}

\subsection{Basis}

In the {\it ab initio} NCSM, we use a HO basis. A single-nucleon HO wave function can be written 
as
 \begin{equation}\label{HOwave}
\varphi_{nlm}(\vec{r};b)=R_{nl}(r;b)Y_{lm}(\hat{r})  \; ,
\end{equation}
with $R_{nl}(r,b)$ the radial HO wave function and $b$ the HO length parameter 
related to the HO frequency 
$\Omega$ as $b=\sqrt{\frac{\hbar}{m\Omega}}$, with $m$ the nucleon mass. The HO length 
parameter $b$ is often
dropped in $R_{nl}$ and $\varphi_{nlm}$ in the following text to simplify notation.

The HO wave functions have important transformation properties that we utilize frequently.
Let us consider two particles with different masses of ratio $d=m_2/m_1$ moving in a HO well.
Their relative and center-of-mass coordinates can be defined by an orthogonal 
transformation 
\begin{eqnarray}\label{coord_tr}
\vec{r}&=&\sqrt{\frac{d}{1+d}}\vec{r}_1-\sqrt{\frac{1}{1+d}}\vec{r}_2 \; ,\\
\vec{R}&=&\sqrt{\frac{1}{1+d}}\vec{r}_1+\sqrt{\frac{d}{1+d}}\vec{r}_2  \; ,
\end{eqnarray}
where all the vectors in (\ref{coord_tr}) and (5)
are defined as products of square root of the respective
mass and the position vector: $\vec{r}=\sqrt{m}\vec{x}$.
The product of the single-particle HO wave functions can then be expressed as a linear combination
of the relative-coordinate HO wave function and the CM coordinate HO wave function:
\begin{equation}\label{HOtransform}
[\varphi_{n_1l_1}(\vec{r}_1)\varphi_{n_2l_2}(\vec{r}_2)]^{(K)}_k=\sum_{nlNL}\langle nl NL K 
| n_1 l_1 n_2 l_2 K\rangle_d [\varphi_{nl}(\vec{r})\varphi_{NL}(\vec{R})]^{(K)}_k \; ,
\end{equation}
with $\langle nl NL K | n_1 l_1 n_2 l_2 K\rangle_d$ a generalized HO bracket that can be evaluated
according to the algorithm in, e.g. Ref. \cite{Tr72}.

As the NN and NNN interactions depend on relative coordinates and/or momenta, the
natural coordinates in the nuclear problem are the relative, or Jacobi, coordinates.

We work in the isospin formalism and consider nucleons with the mass
$m$. A generalization to the proton-neutron formalism with
unequal masses for the proton and the neutron is straightforward.
We will use Jacobi coordinates that are introduced as an orthogonal
transformation of the single-nucleon coordinates. In general,
Jacobi coordinates are proportional to differences of centers of mass
of nucleon sub-clusters. 

For the present purposes we consider just a single set
of Jacobi coordinates. More general discussion can be found in Ref.~\cite{Jacobi_NCSM}.
The following set 
%
\begin{eqnarray}\label{jacobiam11}
\vec{\xi}_0 &=& \sqrt{\frac{1}{A}}\left[\vec{r}_1+\vec{r}_2
                                   +\ldots +\vec{r}_A\right]
\; , \\
\vec{\xi}_1 &=& \sqrt{\frac{1}{2}}\left[\vec{r}_1-\vec{r}_2
                                                     \right]
\; , \\
\vec{\xi}_2 &=& \sqrt{\frac{2}{3}}\left[\frac{1}{2}
                 \left(\vec{r}_1+\vec{r}_2\right)
                                   -\vec{r}_3\right]
\; , \\
&\ldots & \nonumber
\\
\vec{\xi}_{A-2} &=& \sqrt{\frac{A-2}{A-1}}\left[\frac{1}{A-2}
      \left(\vec{r}_1+\vec{r}_2 + \ldots+ \vec{r}_{A-2}\right)
                                   -\vec{r}_{A-1}\right]
\; , \\
\vec{\xi}_{A-1} &=& \sqrt{\frac{A-1}{A}}\left[\frac{1}{A-1}
      \left(\vec{r}_1+\vec{r}_2 + \ldots+ \vec{r}_{A-1}\right)
                                   -\vec{r}_{A}\right]
\; ,
\end{eqnarray}
%
is useful for the construction of the antisymmetrized HO basis.
Here, $\vec{\xi}_0$ is proportional to the center of mass of the
$A$-nucleon system. On the other hand, $\vec{\xi}_\rho$ is proportional
to the relative position of the $\rho+1$-st nucleon and the
center of mass of the $\rho$ nucleons.  

As nucleons are fermions, we need to construct an antisymmetrized basis.
Here we illustrate how to do this for the simplest case of three nucleons.
One starts by introducing a  
HO basis that depends on Jacobi coordinates
$\vec{\xi}_1$ and $\vec{\xi}_2$, defined in Eqs. (8) and (9), 
e.g.,
\begin{equation}\label{hobas}
|(n l s j t; {\cal N} {\cal L} {\cal J}) J T \rangle \; .
\end{equation}
Here $n, l$ and ${\cal N}, {\cal L}$ are the HO quantum numbers
corresponding to the harmonic oscillators associated with the coordinates 
(and the corresponding momenta) $\vec{\xi}_1$ and $\vec{\xi}_2$, respectively. 
The quantum numbers $s,t,j$ describe the spin, isospin and angular momentum
of the relative-coordinate two-nucleon channel of nucleons 1 and 2, while 
${\cal J}$ is the angular momentum of the third nucleon relative to the
center of mass of nucleons 1 and 2. The $J$ and $T$ are the total angular 
momentum and the total isospin, respectively.
Note that the basis (\ref{hobas}) is antisymmetrized with respect
to the exchanges of nucleons 1 and 2, as the two-nucleon channel
quantum numbers are restricted by the condition $(-1)^{l+s+t}=-1$.
It is not, however, antisymmetrized with respect to the exchanges of nucleons
$1\leftrightarrow 3$ and $2\leftrightarrow 3$.
In order to construct a completely antisymmetrized basis, one needs to 
obtain eigenvectors of the antisymmetrizer 
\begin{equation}\label{antisymm3}
{\cal X}=\frac{1}{3}(1+{\cal T}^{(-)}+{\cal T}^{(+)}) \; ,
\end{equation}
where ${\cal T}^{(+)}$ and ${\cal T}^{(-)}$ are the cyclic and the anti-cyclic 
permutation operators, respectively. The antisymmetrizer ${\cal X}$
is a projector satisfying ${\cal X} {\cal X}={\cal X}$. 
When diagonalized in the basis (\ref{hobas}), its eigenvectors
span two eigenspaces. One, corresponding to the eigenvalue 1, is formed
by physical, completely antisymmetrized states and the other, corresponding
to the eigenvalue 0, is formed by spurious states. There are about 
twice as many spurious states as the physical ones \cite{NBG99}.

Due to the antisymmetry with respect to the exchanges $1\leftrightarrow 2$,
the matrix elements in the basis (\ref{hobas}) of the antisymmetrizer 
${\cal X}$ can be evaluated simply as 
$\langle {\cal X} \rangle = \frac{1}{3}(1-2\langle P_{2,3}\rangle)$,
where $P_{2,3}$ is the transposition operator corresponding to the exchange
of nucleons 2 and 3.
Its matrix element can be evaluated in a straightforward way, e.g.,
\begin{eqnarray}\label{t13t23}
&&\langle (n_1 l_1 s_1 j_1 t_1; {\cal N}_1 {\cal L}_1  
{\cal J}_1) J T | P_{2,3} |  
(n_2 l_2 s_2 j_2 t_2; {\cal N}_2 {\cal L}_2  
{\cal J}_2) J T\rangle 
\nonumber \\
&=& \delta_{N_1,N_2} \hat{t}_1 \hat{t}_2
\left\{ \begin{array}{ccc} \textstyle{\frac{1}{2}} & \textstyle{\frac{1}{2}} 
               & t_1 \\
             \textstyle{\frac{1}{2}}  &  T  & t_2
\end{array}\right\}
\nonumber \\
&& \times \sum_{LS} \hat{L}^2 \hat{S}^2
\hat{j}_1 \hat{j}_2 \hat{\cal J}_1 \hat{\cal J}_2 \hat{s}_1 \hat{s}_2
 (-1)^L
          \left\{ \begin{array}{ccc} l_1   & s_1   & j_1   \\
          {\cal L}_1  & \textstyle{\frac{1}{2}}  & {\cal J}_1 \\
                                     L   & S  & J
\end{array}\right\}
          \left\{ \begin{array}{ccc} l_2   & s_2   & j_2   \\
       {\cal L}_2  & \textstyle{\frac{1}{2}}   & {\cal J}_2 \\
                                     L   & S  & J
\end{array}\right\}
\nonumber \\
&& \times
\left\{ \begin{array}{ccc} \textstyle{\frac{1}{2}} & \textstyle{\frac{1}{2}} 
               & s_1 \\
             \textstyle{\frac{1}{2}}  &  S  & s_2
\end{array}\right\}
\langle n_1 l_1 {\cal N}_1 {\cal L}_1 L 
| {\cal N}_2 {\cal L}_2 n_2 l_2 L \rangle_{\rm 3} \; ,
\end{eqnarray}
where $N_i=2n_i+l_i+2{\cal N}_i+{\cal L}_i, i= 1,2$; 
$\hat{j}=\sqrt{2j+1}$; 
and $\langle n_1 l_1 {\cal N}_1 {\cal L}_1  L 
| {\cal N}_2 {\cal L}_2 n_2 l_2 L \rangle_{\rm 3}$
is the general HO bracket for two particles with mass 
ratio 3 as defined, e.g.,
in Ref. \cite{Tr72}.
The expression (\ref{t13t23}) can be derived by examining 
the action
of $P_{2,3}$ on the basis states (\ref{hobas}).
That operator changes the state 
$|nl(\vec{\xi}_1),{\cal NL}(\vec{\xi}_2), L\rangle$
to $|nl(\vec{\xi'}_1),{\cal NL}(\vec{\xi'}_2), L\rangle$,
where $\vec{\xi'}_i, i=1,2$ are defined as $\vec{\xi}_i, i=1,2$
but with the single-nucleon indexes 2 and 3 exchanged. The primed Jacobi 
coordinates can be expressed as an orthogonal transformation
of the unprimed ones, see Eq. (\ref{HOtransform}). Consequently, 
the HO wave functions
depending on the primed Jacobi coordinates can be expressed
as an orthogonal transformation of the original HO wave functions.
Elements of the transformation are the generalized HO brackets
for two particles with the mass ratio $d$, with $d$ determined
from the orthogonal transformation of the coordinates, see Eq. (\ref{coord_tr}).

The resulting antisymmetrized states can be classified 
and expanded in terms of the original basis (\ref{hobas}) as follows
\begin{equation}\label{abas3}
|N i J T\rangle = \sum \langle nlsjt; {\cal NLJ}||N i J T\rangle 
|(nlsjt;{\cal NLJ}) JT\rangle \; ,
\end{equation}
where $N=2n+l+2{\cal N}+{\cal L}$ and 
where we introduced an
additional quantum number $i$ that distinguishes states with
the same set of quantum numbers $N, J, T$, e.g.,
$i=1,2, \ldots r$ with 
$r$ the total number of antisymmetrized states for a given $N, J, T$.
The symbol $\langle nlsjt; {\cal NLJ}||N i J T\rangle$ is a coefficient
of fractional parentage.

A generalization to systems of more than three nucleons can be done as shown
e.g. in Ref.~\cite{Jacobi_NCSM}. It is obvious, however, that as we increase the number 
of nucleons, the antisymmetrization becomes more and more involved.
Consequently, in the standard shell model calculations one utilizes antisymmetrized
wave functions constructed in a straightforward way as Slater determinants of 
single-nucleon wave functions depending on single-nucleon coordinates $\varphi_i(\vec{r}_i)$. 
It follows from the transformations (\ref{HOtransform}) that a use of a Slater determinant basis 
constructed from single nucleon HO wave functions such as
\begin{equation}\label{HOwavesp}
\varphi_{nljm m_t}(\vec{r},\sigma,\tau;b)=R_{nl}(r;b)
(Y_{l}(\hat{r})\chi(\sigma))^{(j)}_m\chi(\tau)_{m_t}  \; ,
\end{equation}
results in eigenstates of a translationally invariant Hamiltonian that factorize
as products of a wave function depending on relative coordinates and a wave function 
depending on the CM
coordinates. This is true as long as the basis truncation is done by a chosen maximum
of the sum of all HO excitations, i.e.: $\sum_{i=1}^A(2n_i+l_i)\leq N_{totmax}$.
In Eq.~(\ref{HOwavesp}), $\sigma$ and $\tau$ are spin and isospin coordinates of the nucleon.
The physical eigenstates of an translationally invariant Hamiltonian can then be selected
as eigenstates with the CM in the $0\hbar\Omega$ state:
\begin{eqnarray}\label{state_relation}
&&\langle \vec{r}_1 \ldots \vec{r}_A \sigma_1 \ldots \sigma_A \tau_1 \ldots \tau_A 
| A \lambda J M T M_T\rangle_{\rm SD} 
\nonumber \\
&=& 
\langle \vec{\xi}_1 \ldots \vec{\xi}_{A-1}
\sigma_1 \ldots \sigma_A \tau_1 \ldots \tau_A 
| A \lambda J M T M_T\rangle \varphi_{000}(\vec{\xi}_0;b) \; ,
\end{eqnarray}
For a general single-nucleon wave function this factorization is not possible. 
A use of any other 
single nucleon wave function than the HO wave function will result in mixing 
of CM and internal motion.

In the {\it ab initio} NCSM calculations, we use both the Jacobi-coordinate HO basis
and the single-nucleon Slater determinant HO basis. One can choose, whichever
is more convenient for the problem to be solved. One can also mix the two types of bases.
In general for systems of $A\leq4$, the Jacobi coordinate basis is more efficient as one can
perform the antisymmetrization easily. The CM degrees of freedom can be explicitly removed
and a coupled $J^\pi T$ basis can be utilized with matrix dimensions of the order of thousands.
For systems with $A>4$, it is in general more efficient to use the Slater determinant
HO basis. In fact, we use so-called m-scheme basis with conserved 
quantum numbers $M=\sum_{i=1}^Am_i$,
parity $\pi$ and $M_T=\sum_{i=1}^A m_{ti}$. The antisymmetrization is trivial, but the dimensions
can be huge as the CM degrees of freedom are present and no $J T$ coupling is considered.
The advantage is the possibility to utilize the powerful second quantization technique,
shell model codes, transition density codes and so on.

As mentioned above, the model space truncation is always done using the condition
$\sum_{i=1}^A(2n_i+l_i)\leq N_{totmax}$. Often, instead of $N_{totmax}$ we introduce
the parameter $N_{\rm max}$ that measures the maximal allowed HO excitation energy above
the unperturbed ground state. For $A=3,4$ systems $N_{\rm max}=N_{totmax}$. For the $p$-shell nuclei
they differ, e.g. for $^6$Li,  $N_{\rm max}=N_{totmax}-2$, for $^{12}$C, $N_{\rm max}=N_{totmax}-8$ etc.

\subsection{Effective interaction}

In the {\it ab initio} NCSM calculations we use a truncated HO basis
as discussed in previous sections. The inter-nucleon interactions
act, however, in the full space. In order to obtain meaningful results
in the truncated space, or model space, the inter-nucleon interactions
needs to be renormalized. We need to construct an effective Hamiltonian 
with the inter-nucleon interactions replaced by effective interactions.
By meaningful results we understand results as close as possible to the 
full space exact results for a subset of eigenstates. Mathematically 
we can construct an effective Hamiltonian that exactly reproduces
the full space results for a subset of eigenstates. In practice, we cannot in general
construct this exact effective Hamiltonian for the $A$-nucleon problem
we want to solve. However, we can construct an effective Hamiltonian that is exact
for a two-nucleon system or for a three-nucleon system or even for a four-nucleon system.
The corresponding effective interactions can then be used in the $A$-nucleon calculations.
Their use in general improves the convergence of the problem to the exact full space result
with the increase of the basis size. By construction, these effective interactions
converge to the full-space inter-nucleon interactions therefore guaranteeing
convergence to exact solution when the basis size approaches the infinite full space.

In our approach we employ the so-called Lee-Suzuki similarity transformation
method \cite{LS80,S82SO83}, which yields 
a starting-energy independent hermitian effective interaction.
We first recapitulate general formulation and basic results
of this method. Applications of this method for computation of two- or
three-body effective interactions are described afterwards. 

Let us consider an ${\it arbitrary}$ Hamiltonian $H$ with the eigensystem
$E_k, |k\rangle$, i.e.,
\begin{equation}\label{schreq}
H|k\rangle = E_k |k\rangle \; .
\end{equation}
Let us further divide the full space into the model space defined by 
a projector $P$ and the complementary space defined by a projector
$Q$, $P+Q=1$.
A similarity transformation of the Hamiltonian $e^{-\omega} H e^\omega$ 
can be introduced with a transformation operator $\omega$ satisfying 
the condition $\omega= Q \omega P$. The transformation operator
is then determined from the requirement of decoupling of the Q-space and
the model space as follows
\begin{equation}\label{decoupl}
Q e^{-\omega} H e^\omega P = 0 \; .
\end{equation}
If we denote the model space basis states as $|\alpha_P\rangle$,
and those which belong to the Q-space, as $|\alpha_Q\rangle$,
then the relation $Q e^{-\omega} H e^\omega P |k\rangle = 0$,
following from Eq. (\ref{decoupl}), will be satisfied for a particular
eigenvector $|k\rangle$ of the Hamiltonian (\ref{schreq}),
if its Q-space components can be expressed as a combination
of its P-space components with the help of the transformation operator 
$\omega$, i.e.,
\begin{equation}\label{eigomega}  
\langle\alpha_Q|k\rangle=\sum_{\alpha_P}
\langle\alpha_Q|\omega|\alpha_P\rangle \langle\alpha_P|k\rangle \; .
\end{equation}
If the dimension of the model space is $d_P$, we may choose a set
${\cal K}$ of $d_P$ eigenevectors, 
for which the relation (\ref{eigomega}) 
will be satisfied. Under the condition that the $d_P\times d_P$ 
matrix defined by matrix elements $\langle\alpha_P|k\rangle$ for $|k\rangle\in{\cal K}$
is invertible, the operator $\omega$ can be determined from 
(\ref{eigomega}) as
\begin{equation}\label{omegasol}
\langle\alpha_Q|\omega|\alpha_P\rangle = \sum_{k \in{\cal K}}
\langle\alpha_Q|k\rangle\langle\tilde{k}|\alpha_P\rangle \; ,
\end{equation}  
where we denote by tilde the inverted matrix of $\langle\alpha_P|k\rangle$, e.g.,
$\sum_{\alpha_P}\langle\tilde{k}|\alpha_P\rangle\langle\alpha_P
|k'\rangle = \delta_{k,k'}$, for $k,k'\in{\cal K}$.

The hermitian effective Hamiltonian defined on the model space $P$
is then given by \cite{S82SO83}
\begin{equation}\label{hermeffomega}
\bar{H}_{\rm eff}
=\left[P(1+\omega^\dagger\omega)P\right]^{1/2}
PH(P+Q\omega P)\left[P(1+\omega^\dagger\omega)
P\right]^{-1/2} \; .
\end{equation}
By making use of the properties of the operator $\omega$,
the effective Hamiltonian $\bar{H}_{\rm eff}$ can be rewritten
in an explicitly hermitian form as
\begin{equation}\label{exhermeff}
\bar{H}_{\rm eff}
=\left[P(1+\omega^\dagger\omega)P\right]^{-1/2}
(P+P\omega^\dagger Q)H(Q\omega P+P)\left[P(1+\omega^\dagger\omega)
P\right]^{-1/2} \; .
\end{equation}
With the help of the solution for $\omega$ (\ref{omegasol})
we obtain a simple expression for the matrix elements of 
the effective Hamiltonian
\begin{eqnarray}\label{effham}
\langle \alpha_P | \bar{H}_{\rm eff} |\alpha_{P'}\rangle
&=& \sum_{k \in{\cal K}}\sum_{\alpha_{P''}}\sum_{\alpha_{P'''}} 
\langle \alpha_P | (1+\omega^\dagger\omega)^{-1/2}
|\alpha_{P''}\rangle
\langle \alpha_{P''} |\tilde{k}\rangle E_k 
\langle\tilde{k}|\alpha_{P'''}\rangle
\nonumber \\
&&\times
\langle \alpha_{P'''} | (1+\omega^\dagger\omega)^{-1/2}
|\alpha_{P'}\rangle \; .
\end{eqnarray}
For computation of the matrix elements of 
$(1+\omega^\dagger\omega)^{-1/2}$, we can use the relation
\begin{equation}\label{omdegom}
\langle \alpha_P | (1+\omega^\dagger\omega)
|\alpha_{P''}\rangle = \sum_{k \in{\cal K}} \langle \alpha_P |
\tilde{k} \rangle \langle \tilde{k}|\alpha_{P''}\rangle \; ,
\end{equation}
to remove the summation over the Q-space basis states.
The effective Hamiltonian (\ref{effham}) reproduces the
eigenenergies $E_k, k\in{\cal K}$ in the model space.

It has been shown \cite{UMOA} that the hermitian effective 
Hamiltonian (\ref{exhermeff}) can be obtained
directly by a unitary transformation of the original Hamiltonian:
\begin{equation}\label{UMOAtrans}
\bar{H}_{\rm eff} = Pe^{-S} H e^{S}P \; ,
\end{equation}
with an anti-hermitian operator $S={\rm arctanh}(\omega-\omega^\dagger)$.
The transformed Hamiltonian then satisfies decoupling conditions
$Qe^{-S} H e^{S}P=Pe^{-S} H e^{S}Q=0$.

We can see from Eqs.~(\ref{effham}) and (\ref{omdegom}) that in order
to construct the effective Hamiltonian we need to know a subset of exact
eigenvalues and model space projections of a subset of exact eigenvectors.
This may suggest that the method is rather impractical. Also, it follows
from Eq.~(\ref{effham}) that the effective Hamiltonian contains many-body terms,
in fact for an $A$-nucleon system all terms up to $A$-body will in general 
appear in the effective Hamiltonian even if the original Hamiltonian
consisted of just two-body or two- plus three-body terms.

In the {\it ab initio} NCSM we use the above effective interaction theory
as follows. Since the two-body part dominates the $A$-nucleon 
Hamiltonian (\ref{hamomega}), it is reasonable to expect that a two-body
effective interaction that takes into account full space two-nucleon correlations
would be the most important part of the exact effective interaction. If the NNN
interaction is taken into account, a three-body effective interaction that
takes into account full space three-nucleon correlations would be a good
approximation to the exact $A$-body effective interaction. We construct
the two-body or three-body effective interaction by application of 
the above described Lee-Suzuki procedure to a two-nucleon or three-nucleon system.
The resulting effective interaction is then exact for the two- or three-nucleon
system. It is an approximation of the exact $A$-nucleon effective interaction.

Using the notation of Eq.(\ref{hamomega}), the two-nucleon effective interaction is obtained as
\begin{equation}\label{V2eff}
V_{\rm 2eff,12} = P_2[e^{-S_{12}}(h_1+h_2+V^{\Omega,A}_{12})e^{S_{12}}-(h_1+h_2)]P_2   \; ,
\end{equation}
with $S_{12}={\rm arctanh}(\omega_{12}-\omega_{12}^\dagger)$ and $P_2$ is a two-nucleon
model space projector. The two-nucleon model space is defined by a truncation $N_{\rm 12max}$
corresponding to the $A$-nucleon $N_{\rm max}$. For example, for $A=3,4$, $N_{\rm 12max}=N_{\rm max}$,
for $p$-shell nuclei with $A>5$ $N_{\rm 12max}=N_{\rm max}+2$. The operator $\omega_{12}$
is obtained with the help of Eq. (\ref{omegasol}) from exact solutions of the Hamiltonian
$h_1+h_2+V^{\Omega,A}_{12}$ which are straightforward to find. 
In practice, we actually do not need to calculate
$\omega_{12}$, rather we apply Eqs.~(\ref{effham}) and (\ref{omdegom}) with the two-nucleon
solutions to directly calculate $P_2e^{-S_{12}}(h_1+h_2+V^{\Omega,A}_{12})e^{S_{12}}P_2$. 
To be explicit, the two-nucleon calculation is done with 
\begin{equation}\label{hamomega2}
H_2^\Omega = H_{02}+V^{\Omega,A}_{12}=
\frac{\vec{p}^2}{2m}
+\frac{1}{2}m\Omega^2 \vec{r}^2
+ V_{NN}(\sqrt{2}\vec{r})-\frac{m\Omega^2}{A}\vec{r}^2 \; ,
\end{equation}
where $\vec{r}=\sqrt{\frac{1}{2}}(\vec{r}_1-\vec{r}_2)$ and
$\vec{p}=\sqrt{\frac{1}{2}}(\vec{p}_1-\vec{p}_2)$ and
where $H_{02}$ differs from $h_1+h_2$ by the omission of 
the center-of-mass HO term of nucleons 1 and 2. Since $V^{\Omega,A}_{12}$ acts 
on relative coordinate, the $S_{12}$ is independent of the two-nucleon 
center of mass and the two-nucleon center-of-mass Hamiltonian cancels
out in Eq.~(\ref{V2eff}). We can see that for $A>2$ the solutions of
(\ref{hamomega2}) are bound. The relative-coordinate two-nucleon 
HO states used in the calculation are characterized by
quantum numbers $|nlsjt\rangle$ with the radial and orbital
HO quantum numbers corresponding to coordinate $\vec{r}$ 
and momentum $\vec{p}$. Typically, we solve the two-nucleon
Hamiltonian (\ref{hamomega2}) for all two-nucleon channels up
to $j=8$. For the channels with higher $j$ only the kinetic-energy 
term is used in the many-nucleon calculation.
The model space $P_2$ is defined by the maximal number of allowed HO 
excitations $N_{\rm 12max}$ from the condition $2n+l\leq N_{\rm 12max}$.
In order to construct the operator $\omega$ (\ref{omegasol})
we need to select the set of eigenvectors ${\cal K}$.
We select the lowest states
obtained in each channel. It turns out that these states also have
the largest overlap with the $P_2$ model space. Their number is given 
by the number of basis states satisfying $2n+l\leq N_{\rm 12max}$. 

An improvement over the two-body effective interaction approximation
is the use of three-body effective interaction that takes into account
the full space three-nucleon correlations. If the NNN interaction is included,
the three-body effective interaction approximation is rather essential
for $A>3$ systems. First, let us consider the case with no NNN interaction.
The three-body effective interaction can be calculated as
\begin{eqnarray}\label{v3eff_2b}
&& V_{{\rm 3eff},123}^{\rm NN}=
\nonumber \\
&& P_3 \left[e^{-S^{\rm NN}_{123}}(h_1+h_2+h_3+V_{12}^{\Omega,A}
+V_{13}^{\Omega,A}+V_{23}^{\Omega,A})e^{S^{\rm NN}_{123}}
-(h_1+h_2+h_3)\right] P_3 \; . 
\end{eqnarray}
Here, $S^{\rm NN}_{123}={\rm arctanh}(\omega_{123}-\omega_{123}^\dagger)$ and $P_3$ is a three-nucleon
model space projector. The $P_3$ space contains all three-nucleon states up to the highest 
possible three-nucleon excitation, which can be found in the $P$ space
of the $A$-nucleon system. For example,
for $A=6$ and $N_{\rm max}=6$ ($6\hbar\Omega$) space we have $P_3$ defined 
by $N_{\rm 123max}=8$. Similarly, for the $p$-shell nuclei with $A\geq 7$
and $N_{\rm max}=6$ ($6\hbar\Omega$) space we have $N_{\rm 123max}=9$. 
The operator $\omega_{123}$
is obtained with the help of Eq. (\ref{omegasol}) from exact solutions of the Hamiltonian
$h_1+h_2+h_3+V_{12}^{\Omega,A}+V_{13}^{\Omega,A}+V_{23}^{\Omega,A}$,
which are found using the antisymmetrized three-nucleon Jacobi coordinate HO basis. 
In practice, we again do not need to calculate
$\omega_{123}$, rather we apply Eqs.~(\ref{effham}) and (\ref{omdegom}) with the three-nucleon
solutions. The three-body effective interaction is then used in $A$-nucleon calculations
using the effective Hamiltonian
\begin{equation}\label{Ham_A_Omega_eff_NN}
H^\Omega_{A, {\rm eff}}=\sum_{i=1}^A h_i 
+ \frac{1}{A-2}\sum_{i<j<k}^A V_{{\rm 3eff}, ijk}^{\rm NN} \; ,
\end{equation}
where the $\frac{1}{A-2}$ factor takes
care of over-counting the contribution from the two-nucleon interaction.

If the NNN interaction is included, we need to calculate in addition to (\ref{v3eff_2b})
the following effective interaction
\begin{eqnarray}\label{v3eff}
V_{{\rm 3eff},123}^{\rm NN+NNN}&=&P_3 \left[ e^{-S^{\rm NN+NNN}_{123}}(h_1+h_2+h_3+V_{12}^{\Omega,A}
+V_{13}^{\Omega,A}+V_{23}^{\Omega,A}+V_{{\rm NNN}, 123})e^{S^{\rm NN+NNN}_{123}} \right.
\nonumber \\
&&\left. -(h_1+h_2+h_3)\right] P_3 \;  .
\end{eqnarray}
This three-body effective interaction is obtained using full space solutions
of the Hamiltonian 
$h_1+h_2+h_3+V_{12}^{\Omega,A}+V_{13}^{\Omega,A}+V_{23}^{\Omega,A}+V_{{\rm NNN}, 123}$.
The three-body effective interaction contribution from the NNN interaction 
we then define as 
\begin{equation}\label{v3eff_3b}
V_{{\rm 3eff},123}^{\rm NNN}\equiv V_{{\rm 3eff},123}^{\rm NN+NNN}
-V_{{\rm 3eff},123}^{\rm NN} \; .
\end{equation}
The effective Hamiltonian used in the $A$-nucleon calculation is then
\begin{equation}\label{Ham_A_Omega_eff}
H^\Omega_{A, {\rm eff}}=\sum_{i=1}^A h_i 
+ \frac{1}{A-2}\sum_{i<j<k}^A V_{{\rm 3eff}, ijk}^{\rm NN}
+\sum_{i<j<k}^A V_{{\rm 3eff}, ijk}^{\rm NNN} \; .
\end{equation}
At this point
we also subtract the $H_{\rm CM}$ and, if the Slater determinant basis is to be used, 
we add the Lawson projection term 
$\beta(H_{\rm CM}-\frac{3}{2}\hbar\Omega)$ to shift the spurious
CM excitations.

It should be noted that all the effective interaction calculations are performed
in the Jacobi coordinate HO basis. As discussed above, two-body effective interaction 
is performed in the $|nlsjt\rangle$ basis and the three-body effective interaction 
in the $|N i J T\rangle$ basis (\ref{abas3}). In order to perform the $A$-nucleon
calculation in the Slater determinant HO basis as is typically done for $A>4$,
the effective interaction needs to be transformed to single-nucleon HO basis. 
This is done with help of the HO wave function transformations (\ref{HOtransform}).
The details for the three-body case in particular are given in Refs.~\cite{NO03}
and \cite{Nogga06}. 

As a final remark, we note that the unitary transformation performed on the Hamiltonian
should be also applied to other operators that are used to calculate observables.
If this is done, the model-space-size convergence of observables improves. More details
on calculation of effective operators are given in Ref.~\cite{Ionel}.

\subsection{Convergence tests}

\begin{figure}[hbtp]
\begin{center}
  \includegraphics*[width=0.8\columnwidth]
   {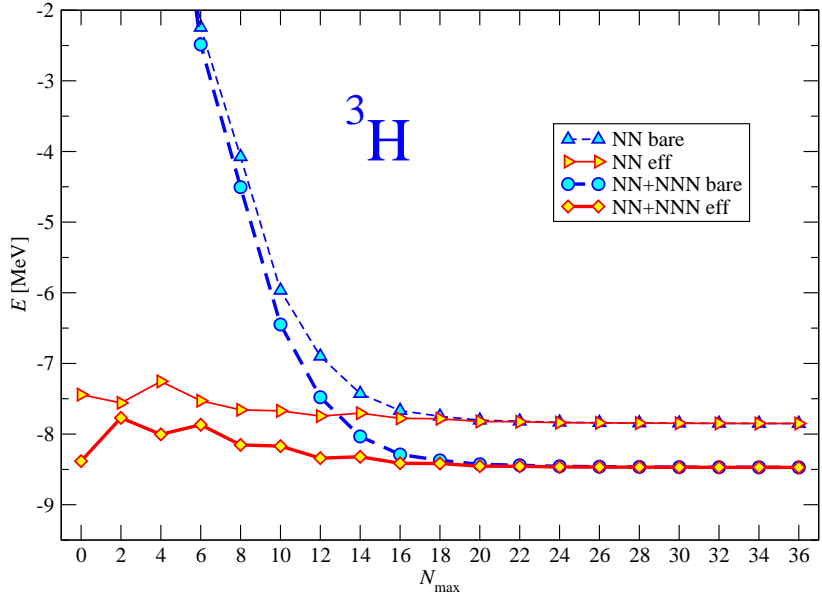}
  \caption{$^3$H ground-state energy dependence on the size of the basis.
The HO frequency of $\hbar\Omega=28$ MeV was employed. Results with (thick lines)
and without (thin lines) the NNN interaction are shown. The full lines correspond 
to calculations with two-body effective interaction derived from the chiral NN interaction, 
the dashed lines to calculations with the bare chiral NN interaction. 
  \label{gs_H3}}
\end{center}
\end{figure}

In this subsection, we give examples of convergence of {\it ab initio} NCSM calculations.
In Fig.~\ref{gs_H3}, we show the
convergence of the $^3$H ground-state energy
with the size of the basis. Thin lines correspond to results obtained with 
the NN interaction only. Thick lines correspond to calculations that also include the 
NNN interaction. The full lines correspond to calculations with two-body effective
interaction derived from the chiral effective field theory (EFT) NN 
interaction of Ref.~\cite{N3LO} discussed in more details
in the next section. The dashed lines correspond 
to calculations with the bare, that is the original unrenormalized, chiral EFT NN interaction. 
The bare NNN interaction is added to either the bare NN or to the effective NN interaction
in calculations depicted by thick lines. Here, we use the chiral EFT NNN interaction that
will be discussed in details in the next section.
We observe that the convergence is faster when
the two-body effective interaction is used. However, starting at about $N_{\rm max}=24$
the convergence is reached also in calculations with the bare NN interaction. The rate
of convergence also depends on the choice of the HO frequency. In general, it is
always advantageous to use the effective interaction in order to improve the convergence
rate. It should be noted that in calculations with the effective interaction, 
the effective Hamiltonian is different at each point as the effective interaction depends
on the size of the model space given by $N_{\rm max}$. The calculation with the bare interaction
is a variational calculation converging from above with $N_{\rm max}$ 
and HO frequency $\Omega$ as variational parameters. The calculation with the effective
interaction is not variational. The convergence can be from above, from below or oscillatory.
This is because a part of the exact effective Hamiltonian is omitted. The calculation
without NNN interaction converges to the $^3$H ground-state energy $-7.852(5)$~MeV, well above
the experimental $-8.482$~MeV. Once the NNN interaction is added, we obtain $-8.473(5)$~MeV,
close to experiment. As discussed in the next section, the NNN parameters were tuned to
reproduce $^3$H and $^3$He binding energies. 

\begin{figure}[hbtp]
\begin{center}
  \includegraphics*[width=0.8\columnwidth]
   {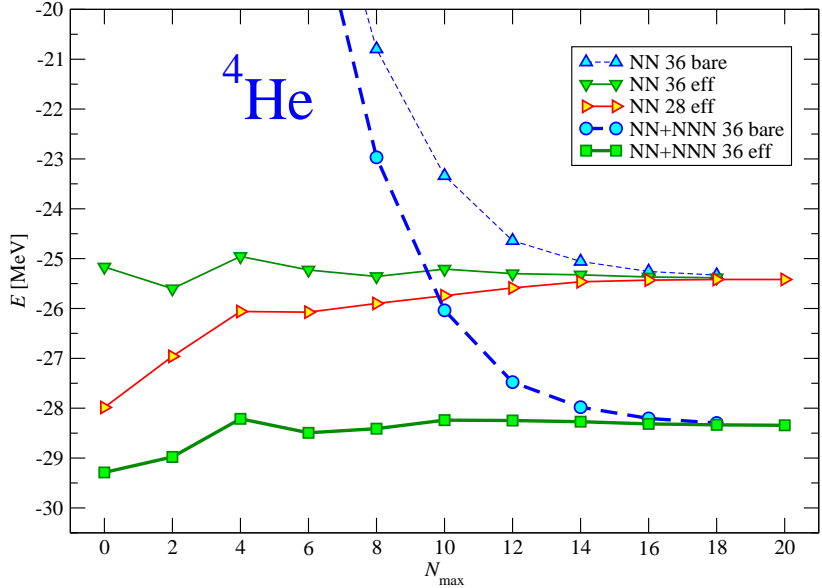}
  \caption{$^4$He ground-state energy dependence on the size of the basis.
The HO frequencies of $\hbar\Omega=28$ and 36 MeV was employed. Results with (thick lines)
and without (thin lines) the NNN interaction are shown. The full lines correspond 
to calculations with three-body effective interaction, 
the dashed lines to calculations with the bare interaction. For further details 
see the text.
  \label{gs_He4}}
\end{center}
\end{figure}

In Fig.~\ref{gs_He4}, we show convergence of the $^4$He ground-state
energy. The NCSM calculations are performed 
in basis spaces up to $N_{\rm max}=20$. Thin lines correspond to results obtained with 
the NN interaction only, while thick lines correspond to calculations that also include the 
NNN interaction. The dashed lines correspond to results obtained with bare interactions.
The full lines correspond to results obtained using three-body effective interaction
(the NCSM three-body cluster approximation). It is apparent that the use of the three-body
effective interaction improves the convergence rate dramatically. We can see that at 
about $N_{\rm max}=18$ the bare interaction calculation reaches convergence as well. 
It should be noted, however, that $p$-shell calculations with the NNN interactions
are presently feasible in model spaces up to $N_{\rm max}=6$ or $N_{\max}=8$. 
The use of the three-body effective interaction is then essential in the $p$-shell 
calculations. The calculation without NNN interaction was done for two different HO
frequencies. it is apparent that convergence to the same result is in both cases.
We note that in the case of no NNN interaction,
we may use just the two-body effective interaction (two-body cluster approximation), which
is much simpler. The convergence is slower, however, see discussion in Ref.~\cite{NO02}.
We also note that $^4$He properties with the chiral EFT NN interaction that we employ
here were calculated using two-body cluster approximation in Ref.~\cite{NC04} and
present results are in agreement with results found there.
Our $^4$He results ground state energy results are $-25.39(1)$~MeV in the NN case 
and $-28.34(2)$~MeV in the NN+NNN case. The experimental value is $-28.296$~MeV. We note that
the present {\it ab initio} NCSM $^3$H and $^4$He results obtained with the chiral EFT
NN interaction are in a perfect agreement with results obtained using the variational
calculations in the hyperspherical harmonics basis as well as with the Faddeev-Yakubovsky
calculations published in Ref.~\cite{HHnonloc}. A satisfying feature of the present
NCSM calculation is the fact that the rate of convergence is not affected
in any significant way by inclusion of the NNN interaction.

\begin{figure}
\begin{center}
\includegraphics[width=1.0\columnwidth]{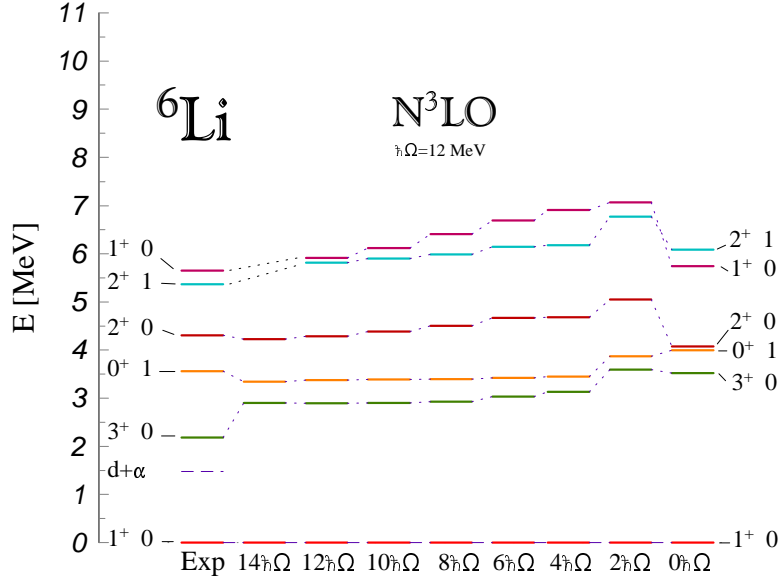}
\caption{\label{li6_exc_12}
Calculated positive-parity excitation spectra of
$^{6}$Li obtained in $0\hbar\Omega$-$14\hbar\Omega$ basis spaces using two-body effective
interactions derived from the chiral EFT NN potential
are compared to experiment. The HO frequency of $\hbar\Omega=12$ MeV was used.
}
\end{center}
\end{figure}

As yet another example of convergence of {\it ab initio} NCSM calculations,
we present the excitation energy calculations of the five lowest excited states of $^6$Li
using the chiral EFT NN potential. The NCSM excitation energy dependence on the basis size
is presented in Fig.~\ref{li6_exc_12} for the HO frequency of $\hbar\Omega=12$ MeV. 
Due to the complexity of the calculations, the lowest four state 
were obtained in basis spaces up to $N_{\rm max}=14$ while we stopped at $N_{\rm max}=12$
for the $2^+ 1$ and the $1^+_2 0$ state.
The calculations were performed using the two-body effective interaction in the Slater 
determinant HO basis with the shell model code Antoine \cite{Antoine}. Results for other
HO frequencies were published in Ref.~\cite{NC04}. 
We observe that the convergence rate with $N_{\rm max}$ is different for different states.
In particular, the $3^+ 0$ state and the $0^+ 1$ state converge faster in the higher frequency 
calculations ($\hbar\Omega=12,13$ MeV), while the higher lying states converge faster 
in the lower frequency calculations ($\hbar\Omega=8,10$ MeV). The results in Fig.~\ref{li6_exc_12}
demonstrate a good convergence of the excitation energies in particular for the $3^+ 0$ and $0^+ 1$
states. An interesting results is the overestimation of the $3^+ 0$ excitation energy
compared to experiment. It turns out that this problem is resolved once the NNN interaction
is included in the Hamiltonian.

\section{Light nuclei from chiral EFT interactions}

Interactions among nucleons are governed by quantum chromodynamics (QCD). 
In the low-energy regime relevant to nuclear structure, 
QCD is non-perturbative, and, therefore, hard to solve. Thus, theory has 
been forced to resort to models for the interaction, which have limited physical basis. 
New theoretical developments, however, allow us connect QCD 
with low-energy nuclear physics. The chiral effective field theory 
($\chi$EFT)~\cite{Weinberg} provides a promising bridge.
Beginning with the pionic or the nucleon-pion system~\cite{bernard95} 
one works consistently with systems of increasing nucleon 
number~\cite{ORK94,Bira,bedaque02a}. 
One makes use of spontaneous breaking of chiral symmetry to systematically 
expand the strong interaction in terms of a generic small momentum
and takes the explicit breaking of chiral symmetry into account by expanding 
in the pion mass. Thereby, the NN interaction, the NNN interaction 
and also $\pi$N scattering are related to each other. 
The $\chi$EFT predicts, along with the NN interaction 
at the leading order, an NNN interaction at the 3rd order (next-to-next-to-leading 
order or N$^2$LO)~\cite{Weinberg,vanKolck:1994,Epelbaum:2002}, 
and even an NNNN interaction at the 4th order (N$^3$LO)~\cite{Epelbaum06}.
The details of QCD dynamics are contained in parameters, 
low-energy constants (LECs), not fixed by the symmetry. These parameters 
can be constrained by experiment. At present, high-quality NN potentials 
have been determined at order N$^3$LO~\cite{N3LO}. 
A crucial feature of $\chi$EFT is the consistency between 
the NN, NNN and NNNN parts. As a consequence, at N$^2$LO and N$^3$LO, except 
for two LECs, assigned to two NNN diagrams, 
the potential is fully 
constrained by the parameters defining the NN interaction.

We adopt the potentials of the $\chi$EFT at the orders presently available, the NN at N$^3$LO  
of Ref.~\cite{N3LO} and the NNN interaction at N$^2$LO \cite{vanKolck:1994,Epelbaum:2002}.
Since the NN interaction is non-local, the {\it ab initio} NCSM
is the only approach currently available 
to solve the resulting many-body Schr\"odinger equation for mid-$p$-shell nuclei.
We are in a position to use the {\it ab initio} NCSM calculations in two
ways. One of them is the determination of the LECs assigned to two NNN diagrams
that must be determined in $A\geq 3$ systems. The other is testing predictions of the
chiral NN and NNN interactions for light nuclei.

\begin{figure}[hbtp]
\vspace{5mm}
\begin{center}
  \includegraphics*[width=0.9\columnwidth]
   {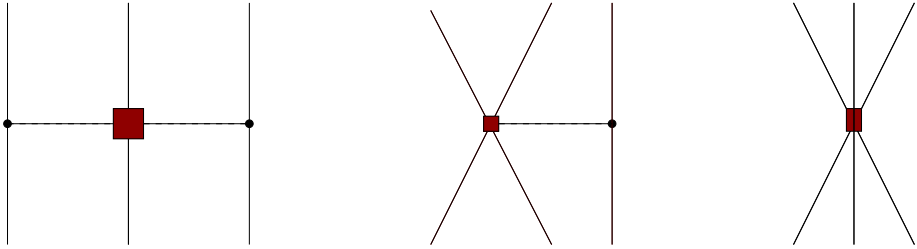}
  \caption{
Terms of the N$^2$LO $\chi$EFT NNN interaction.
  \label{N2LO_NNN}}
\end{center}
\end{figure}

The NNN interaction at N$^2$LO of the $\chi$EFT comprises
of three parts: (i) The two-pion exchange, (ii) the one-pion exchange plus contact
and the three-nucleon contact, see Fig.~\ref{N2LO_NNN}. The LECs associated with the 
two-pion exchange also appear in the NN interaction and are therefore determined
in the $A=2$ system. The one-pion exchange plus contact term (D-term)
is associated with the LEC $c_D$ and the three-nucleon contact term (E-term)
is associated with the LEC $c_E$. The $c_D$ and $c_E$ LECs, expected to be of order 
one, can be constrained by the $A=3$ binding energy. We then still need additional observable
to determine the two parameters. Their determination from three-nucleon scattering data
is difficult due to a correlation of the $^3$H binding energy and, e.g. the $nd$ doublet 
scattering length \cite{Epelbaum:2002} and, in general, due to the lack of an in-depth 
three-nucleon scattering phase shift analysis.
We therefore investigate sensitivity of the $A>3$ nuclei properties to the variation 
of the constrained LECs. 

Before presenting results of the {\it ab initio} NCSM calculations with the
$\chi$EFT NN+NNN interactions, most of which were published in Ref.~\cite{NGVON07},
let us discuss briefly a few technical details of
the calculations with the NNN interaction. 
The NNN interaction is symmetric under permutation of the three nucleon indexes.
It can be written as a sum of three pieces related by particle permutations:
\begin{equation}\label{W}
W=W_1+W_2+W_3   
\end{equation}
To obtain its matrix element in an antisymmetrized three-nucleon basis we need to consider just
a single term, e.g. $W_1$. 
Using the basis introduced in Eq.~(\ref{abas3}), a general matrix element 
can be written as
\begin{eqnarray}\label{mat_el_W}
\langle N i J T | W | N' i^\prime J T \rangle &=& 
3 \langle N i J T | W_1 | N^\prime i^\prime J T \rangle
\nonumber \\
&=& 3 \sum \langle nlsjt, {\cal N L J} || N i J T \rangle 
\langle n^\prime l^\prime s^\prime j^\prime t^\prime, 
{\cal N^\prime L^\prime J^\prime} || N^\prime i^\prime J T \rangle
\nonumber \\
& & 
\times \langle (nlsjt, {\cal N L J})  J T | W_1 | (n^\prime l^\prime s^\prime j^\prime t^\prime, 
{\cal N^\prime L^\prime J^\prime}) J T \rangle   \; .
\end{eqnarray}
We consider just the most trivial part of the $\chi$EFT N$^2$LO NNN interaction, 
the three-nucleon contact term and evaluate its matrix element to demonstrate
that non-trivial effort is needed to include the NNN interactions in many-body calculations.
The three-nucleon contact term can be written as 
\begin{eqnarray}\label{W1_cont}
W_1^{\rm cont} &=& E \vec{\tau}_2 \cdot \vec{\tau}_3 
\delta(\vec{r}_1-\vec{r}_2) \delta(\vec{r}_3-\vec{r}_1)
\nonumber \\
&=& 
E \vec{\tau}_2 \cdot \vec{\tau}_3 \frac{1}{(2\pi)^6} \frac{1}{(\sqrt{3})^3}
\int {\rm d}\vec{\pi}_1 {\rm d}\vec{\pi}_2 {\rm d}\vec{\pi}_1^\prime {\rm d}\vec{\pi}_2^\prime
| \vec{\pi}_1 \vec{\pi}_2 \rangle \langle \vec{\pi}_1^\prime \vec{\pi}_2^\prime | \; ,
\end{eqnarray}
with $E=\frac{c_E}{F_\pi^4\Lambda_\chi}$ where $\Lambda_\chi$ is the chiral symmetry breaking scale 
of the order of the $\rho$ meson mass and $F_\pi=92.4$~MeV is the weak pion decay constant. 
The $\vec{\pi}_1$ and $\vec{\pi}_2$ are Jacobi momenta associated with the 
the Jacobi coordinates $\vec{\xi}_1$ and $\vec{\xi}_2$ defined in Eq.~(8) and (9). 

The contact term must be regulated before it can be used in many-body calculations.
We consider a regulator depending on momentum transfer:
\begin{eqnarray}\label{W1_cont_mt}
W_1^{\rm cont,Q} &=& E \vec{\tau}_2 \cdot \vec{\tau}_3 \frac{1}{(2\pi)^6} \frac{1}{(\sqrt{3})^3}
\int {\rm d}\vec{\pi}_1 {\rm d}\vec{\pi}_2 {\rm d}\vec{\pi}_1^\prime {\rm d}\vec{\pi}_2^\prime
| \vec{\pi}_1 \vec{\pi}_2 \rangle F(\vec{Q}^2;\Lambda) 
F(\vec{Q}^{\prime 2};\Lambda)
\langle \vec{\pi}_1^\prime \vec{\pi}_2^\prime | 
\nonumber \\
&=& E \vec{\tau}_2 \cdot \vec{\tau}_3 \int {\rm d}\vec{\xi}_1 {\rm d}\vec{\xi}_2 
| \vec{\xi}_1 \vec{\xi}_2 \rangle Z_0(\textstyle{\sqrt{2}}\xi_1;\Lambda) 
Z_0(|\textstyle{\frac{1}{\sqrt{2}}}\vec{\xi}_1+\textstyle{\sqrt{\frac{3}{2}}}\vec{\xi}_2|;\Lambda)
\langle \vec{\xi}_1 \vec{\xi}_2 | \; , 
\end{eqnarray}
with the regulator function $F(q^2;\Lambda)= \exp(-q^4/\Lambda^4)$. We defined momenta transferred 
by nucleon 2 and nucleon 3: 
$\vec{Q} = \vec{p}_2^\prime - \vec{p}_2 = -\frac{1}{\sqrt{2}}(\vec{\pi}_1^\prime-\vec{\pi}_1)
+\frac{1}{\sqrt{6}}(\vec{\pi}_2^\prime-\vec{\pi}_2)$
and
$\vec{Q}^\prime = \vec{p}_3^\prime - \vec{p}_3 = -\sqrt{\frac{2}{3}} (\vec{\pi}_2^\prime-\vec{\pi}_2)$. 
Also, we introduced the function
\begin{equation}\label{Z_0}
Z_0(r;\Lambda)= \frac{1}{2\pi^2} \int {\rm d} q q^2 j_0(qr) F(q^2;\Lambda) \;.   
\end{equation}
This results in an interaction local in coordinate space because of the dependence
of the regulator function on differences of initial and final Jacobi momenta.
We can see that after the regulation, the form of the contact interaction
becomes much more complicated. The three-nucleon matrix element of the regulated
term is then obtained in the form
\begin{eqnarray}\label{mat_el_W1_cont_mt}
&&\langle (nlsjt, {\cal N L J})  J T | W_1^{\rm cont,Q} | 
(n^\prime l^\prime s^\prime j^\prime t^\prime, {\cal N^\prime L^\prime J^\prime}) J T \rangle   
\nonumber \\
&=& 
E 6 \delta_{ss^\prime} 
\hat{t}\hat{t}^\prime (-1)^{t+t^\prime+T+\textstyle{\frac{1}{2}}}
\left\{ \begin{array}{ccc} t  & t^\prime  & 1 \\
  \textstyle{\frac{1}{2}} & \textstyle{\frac{1}{2}} & \textstyle{\frac{1}{2}}
\end{array}\right\}
\left\{ \begin{array}{ccc} t  & t^\prime  & 1 \\
  \textstyle{\frac{1}{2}} & \textstyle{\frac{1}{2}} & T
\end{array}\right\} 
\nonumber \\
&&\times
\hat{j}\hat{j}^\prime \hat{\cal J}\hat{\cal J}^\prime \hat{l}^\prime \hat{\cal L}^\prime
(-1)^{J-\textstyle{\frac{1}{2}}+{\cal J}^\prime - {\cal J}+l+{\cal L}+s}
\nonumber \\
&&\times \sum_X (-1)^X \hat{X}^2
\left\{ \begin{array}{ccc} l^\prime & l & X \\
       j & j^\prime & s
\end{array}\right\}
\left\{ \begin{array}{ccc} j & j^\prime & X \\
       {\cal J}^\prime & {\cal J} & J
\end{array}\right\}
\left\{ \begin{array}{ccc} {\cal J}^\prime & {\cal J} & X \\
    {\cal L} & {\cal L}^\prime & \textstyle{\frac{1}{2}} 
\end{array}\right\}
\nonumber \\
&&\times
(l^\prime 0 X 0 | l 0) ({\cal L}^\prime 0 X 0| {\cal L} 0 )
\nonumber \\
&&
\times \int {\rm d}\xi_1 {\rm d}\xi_2
\xi_1^{2} \xi_2^{2}
R_{n l}(\xi_1,\textstyle{b}) 
R_{{\cal N} {\cal L}}(\xi_2,\textstyle{b}) 
R_{n^\prime l^\prime}(\xi_1,\textstyle{b}) 
R_{{\cal N}^\prime {\cal L}^\prime}(\xi_2,\textstyle{b}) 
\nonumber \\
&&\times
Z_0(\textstyle{\sqrt{2}}\xi_1;\Lambda) 
Z_{0,X}(\sqrt{\textstyle{\frac{1}{2}}}\xi_1,\sqrt{\textstyle{\frac{3}{2}}}\xi_2;\Lambda) \;,
\end{eqnarray}
with a new function
%
\begin{equation}\label{Z_0L}
Z_{0,X}(r_1,r_2;\Lambda)= \frac{1}{2\pi^2} \int {\rm d} q q^2 j_X(qr_1) j_X(qr_2) F(q^2;\Lambda)  \; . 
\end{equation}
and customary abbreviation $\hat{l}=\sqrt{2l+1}$.
Evaluation of the other N$^2$LO NNN terms is still more complicated.  

It is important to note that our NCSM results through $A=4$ are fully converged in that 
they are independent of the $N_{\rm max}$ cutoff and the $\hbar\Omega$ HO energy.
This was demonstrated in the subsection on the {\it ab initio} NCSM convergence tests
in particular for the chiral EFT interactions we are investigating here. 
For heavier systems, we characterize the approach to convergence by the dependence 
of results on $N_{\rm max}$ and $\hbar\Omega$.

\begin{figure}
\centerline 
{\includegraphics[width=0.8\columnwidth]{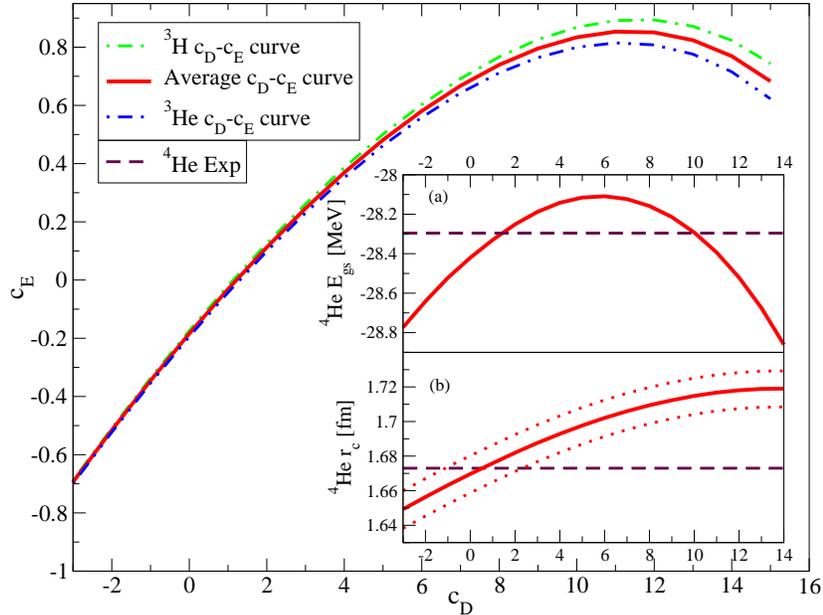}}
\caption{Relations between $c_D$ and $c_E$ for which  the 
binding energy of $^3$H ($8.482$ MeV) and  $^3$He ($7.718$ MeV) are reproduced. 
(a) $^4$He ground-state energy along the averaged curve. 
(b) $^4$He charge radius $r_c$ along the averaged curve. Dotted lines represent 
the $r_c$ uncertainty due to the uncertainties in the proton charge radius.}
\label{CDCE_curve}
\end{figure}

Fig.~\ref{CDCE_curve} shows the trajectories of the two LECs $c_D-c_E$ that 
are determined from fitting the binding energies of the $A=3$ systems. 
Separate curves are shown for $^3$H and $^3$He fits, as well as their average.
There are two points where the binding of $^4$He is reproduced exactly.  
We observe, however, that in the whole investigated range of $c_D-c_E$, the calculated 
$^4$He  binding energy  is within a few hundred keV of experiment. 
Consequently, the determination of the LECs in this way is likely not very stringent.
We therefore investigate the sensitivity of the $p$-shell nuclear properties 
to the choice of the $c_D-c_E$ LECs. First, we maintain the $A=3$ binding energy 
constraint. Second, we limit ourselves to the $c_D$ values in the vicinity 
of the point $c_{D}\sim 1$ since the values close to the point 
$c_{D}\sim 10$ overestimate the $^4$He radius.
Also this large value might be considered ``unnatural'' from the $\chi$EFT 
point of view.

\begin{figure}
\vspace{1cm}
\begin{center}
{\includegraphics[width=0.8\columnwidth]{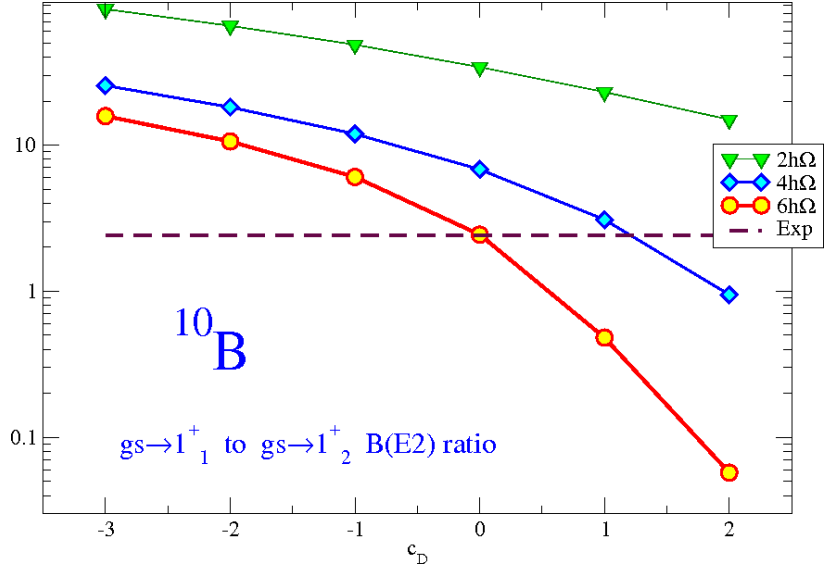}}
\caption{Dependence on the $c_D$ with the $c_E$ constrained by the $A=3$ 
binding energy fit for different basis sizes for $^{10}$B 
B(E2;$3^+_1 0 \rightarrow 1^+_1 0$)/B(E2;$3^+_1 0 \rightarrow 1^+_2 0$) ratio.
The HO frequency of $\hbar\Omega=14$ MeV was employed.}
\label{cD_dep_10B}
\end{center}
\end{figure}

While most of the $p$-shell nuclear properties, e.g. excitation spectra,
are not very sensitive to variations of $c_D$ in the vicinity of the $c_{D}\sim 1$ point,
we were able to identify several observables that do demonstrate strong dependence on $c_D$. 
For example, the $^6$Li quadrupole moment that changes sign 
depending on the choice of $c_D$. In Fig.~\ref{cD_dep_10B}, we display the ratio 
of the B(E2) transitions from the $^{10}$B ground state to the first and the second $1^+ 0$ state.
This ratio changes by several orders of magnitude 
depending on the $c_D$ variation. This is due to the fact
that the structure of the two $1^+ 0$ states is exchanged depending on $c_D$.

\begin{figure}
\vspace{1cm}
\begin{center}
{\includegraphics[width=0.8\columnwidth]{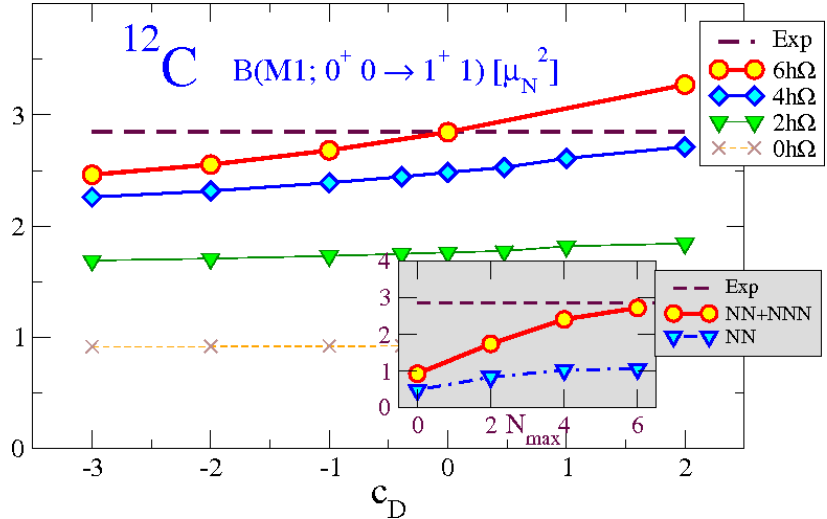}}
\caption{Dependence on the $c_D$ with the $c_E$ constrained by the $A=3$ 
binding energy fit for different basis sizes for 
the $^{12}$C B(M1;$0^+ 0\rightarrow 1^+ 1$). 
The HO frequency of $\hbar\Omega=15$ MeV was employed.
In the inset, the convergence of the 
B(M1;$0^+ 0\rightarrow 1^+ 1$) is presented for calculations with 
(using $c_D=-1$) and without the NNN interaction.}
\label{cD_dep_12C}
\end{center}
\end{figure}

In Fig.~\ref{cD_dep_12C}, we present the $^{12}$C B(M1) transition
from the ground state to the $1^+ 1$ state. The B(M1) transition inset  
illustrates the importance of the NNN interaction in reproducing 
the experimental value \cite{Hayes03}. Overall our results show that 
for $c_D<-2$ the $^4$He radius 
and the $^6$Li quadrupole moment underestimate experiment while for $c_D>0$ the lowest
two $1^+$ states of $^{10}$B are reversed and the $^{12}$C B(M1;$0^+0\rightarrow 1^+1$)
is overestimated. We therefore select $c_D=-1$ 
as globally the best choice and use it for our further investigation. 

\begin{figure}
{\includegraphics[width=0.9\columnwidth]{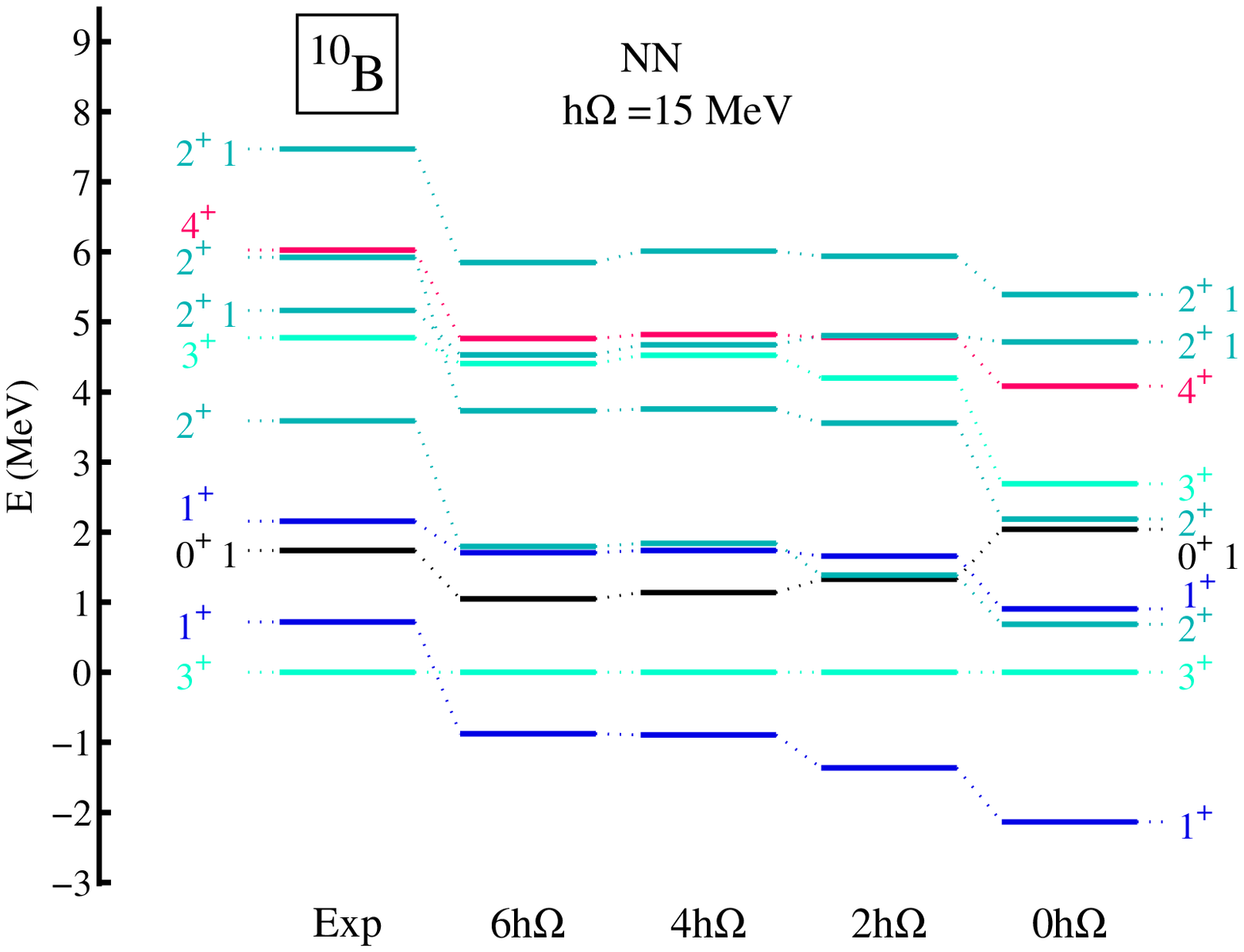}}
\caption{$^{10}$B excitation spectra as function of the basis space size $N_{\rm max}$ 
at $\hbar\Omega=15$ MeV using the chiral NN interaction
and comparison with experiment. The isospin of the states 
not explicitly depicted is $T$=0.}
\label{B10_NN}
\end{figure}

\begin{figure}
{\includegraphics[width=0.9\columnwidth]{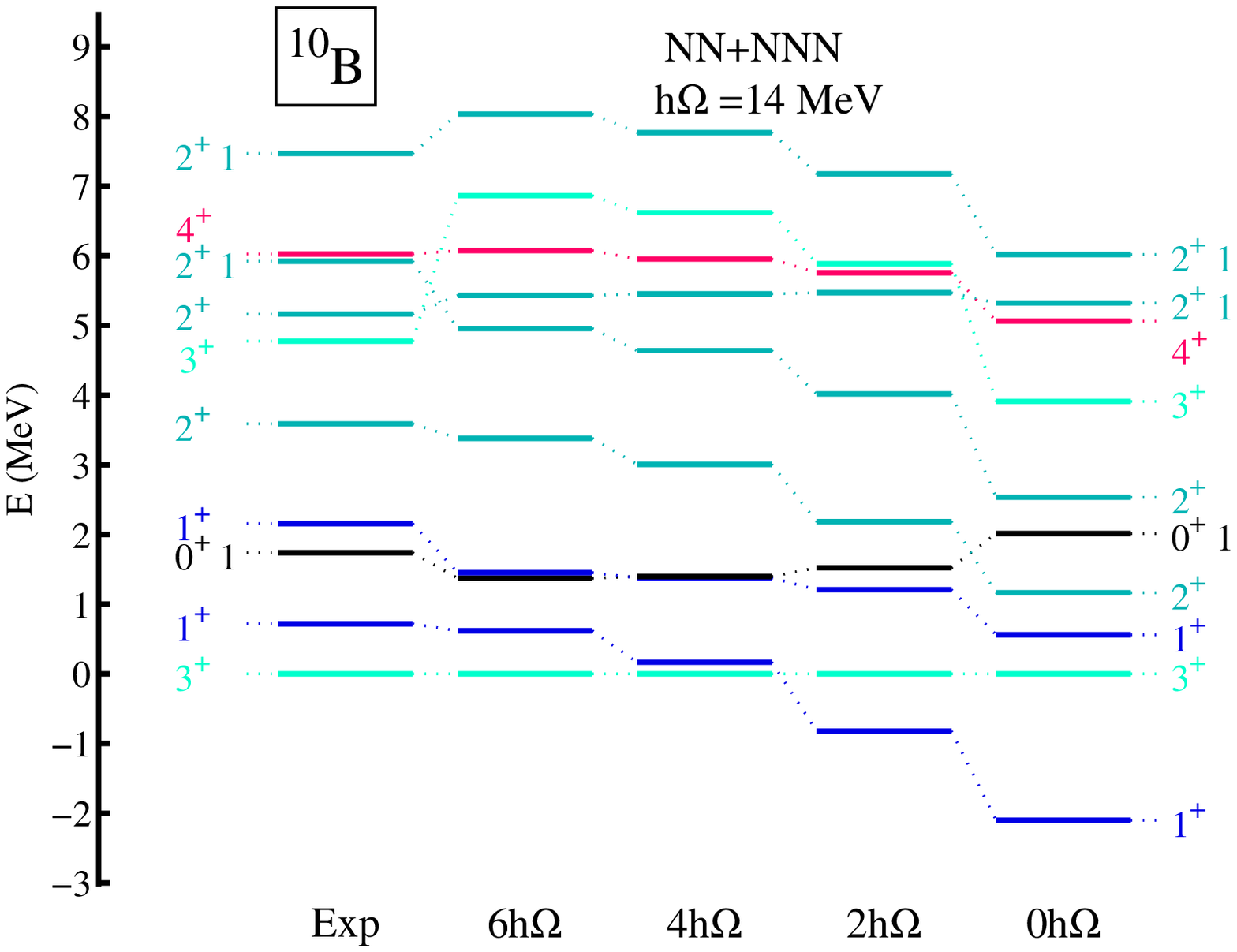}}
\caption{$^{10}$B excitation spectra as function of the basis space size $N_{\rm max}$ 
at $\hbar\Omega=14$ MeV using the chiral NN+NNN interaction
and comparison with experiment. The isospin of the states 
not explicitly depicted is $T$=0.}
\label{B10_NN_NNN}
\end{figure}

We present in Figs.~\ref{B10_NN} and \ref{B10_NN_NNN} the excitation spectra of $^{10}$B as 
a function of $N_{\rm max}$ for both the chiral NN+NNN, as well as with 
the chiral NN interaction alone. In both cases, the convergence 
with increasing $N_{\rm max}$ is quite quite reasonable for the low-lying states. 
Similar convergence rates are obtained for our other $p-$shell nuclei.

A remarkable feature of the $^{10}$B results is the observation that the chiral NN
interaction alone predicts incorrect ground-state spin of $^{10}$B. The experimental
value is $3^+ 0$ while the calculated one is $1^+ 0$. On the other hand, once we also 
include the chiral NNN interaction in the Hamiltonian, which is actually required by 
the $\chi$EFT, the correct ground-state spin is predicted. Further, once we select the $c_D$
value as discussed above, i.e. $c_D=-1$, we also obtain the two lowest $1^+ 0$ states
in the experimental order.  

\begin{figure}[htb]
{\includegraphics[width=1.0\columnwidth]{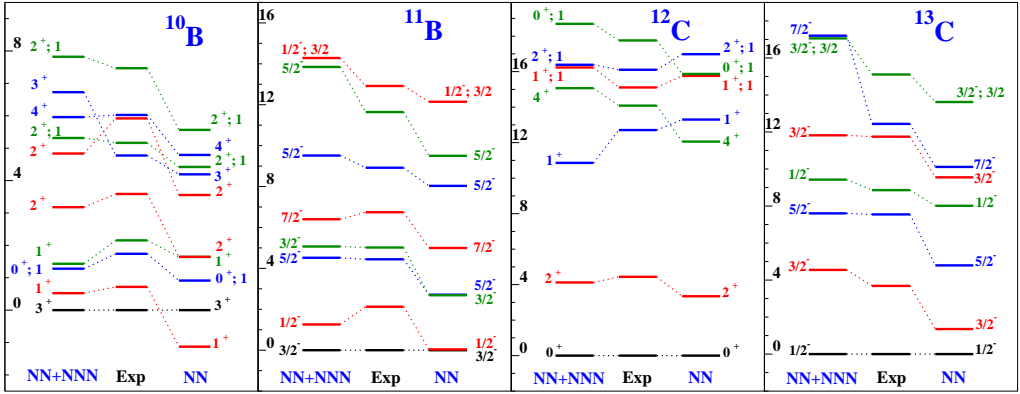}}
\caption{States dominated by $p$-shell configurations for $^{10}$B, $^{11}$B, $^{12}$C, 
and $^{13}$C calculated at $N_{\rm max}=6$ using  $\hbar\Omega=15$ MeV 
(14 MeV for $^{10}$B). Most of the eigenstates are isospin $T$=0 or 1/2, 
the isospin label is explicitly shown only for states with $T$=1 or 3/2.
The excitation energy scales are in MeV.}
\label{B10B11C12C13}
\end{figure}

We display in Fig.~\ref{B10B11C12C13} the natural parity excitation spectra of four nuclei 
in the middle of the $p-$shell with both the NN and the NN+NNN effective interactions 
from $\chi$EFT. The results shown are obtained 
in the largest basis spaces achieved to date for these nuclei with the NNN interactions, 
$N_{\rm max}=6$ ($6\hbar\Omega$). Overall, the NNN interaction contributes 
significantly to improve theory in comparison with experiment. 
This is especially well-demonstrated in the odd mass nuclei for the lowest few excited states. 
The case of the ground state spin of $^{10}$B and its sensitivity to the presence 
of the NNN interaction discussed also in Figs.~\ref{B10_NN} and \ref{B10_NN_NNN}
is clearly evident. We note that the $^{10}$B results with the NN interaction only
in Fig.~\ref{B10_NN}
were obtained with the HO frequency of $\hbar\Omega=15$ MeV, while those
in Fig.~\ref{B10B11C12C13} with $\hbar\Omega=14$ MeV. A weak HO frequency dependence
of the $N_{\rm max}=6$ results is evident. The $^{10}$B results with NN+NNN interaction
presented in Figs.~\ref{B10_NN_NNN}
and \ref{B10B11C12C13} were obtained using the same HO frequency. Still, one may notice
small differences of the $N_{\rm max}=6$ results. The reason behind those differences
is the use of two alternative D-term regularizations (both depending on the momentum transfer,
 details will be discussed elsewhere). 
As the dependence on the regulator is a higher order
effect than the $\chi$EFT expansion order used to derive the NNN interaction, 
these differences should have only minor effect. It is satisfying that the present
$^{10}$B results appear to support this expectation.

Concerning the $^{12}$C results, there is an initial indication 
that the chiral NNN interaction is somewhat over-correcting the inadequacies of the NN interaction 
since, e.g. $1^+ 0$ and the $4^+ 0$ states in $^{12}$C
are not only interchanged but they are also spread apart 
more than the experimentally observed separation. 
In the $^{13}$C results, we can also identify an indication of 
an overly strong correction arising from the chiral NNN interaction
as seen in the upward shift of the $\frac{7}{2}^-$ state.
However, the experimental $\frac{7}{2}^-$ may have significant intruder components 
and is not well-matched with our state. 
In addition, convergence for some higher lying states is affected by incomplete treatment 
of clustering in the NCSM. This point will be elaborated upon later. 
These results required substantial computer resources. 
A typical $N_{\rm max}=6$ spectrum shown in Fig. \ref{B10B11C12C13} 
and a set of additional experimental observables, 
takes 4 hours on 3500 processors of the LLNL's Thunder machine. 
The $A$-nucleon calculations were performed in the Slater determinant
HO basis using the shell model code MFD \cite{MFD}. 
More details on some of the results discussed here are published Ref.~\cite{NGVON07}.

The calculations presented in this section demonstrate that the chiral NNN 
interaction makes substantial contributions 
to improving the spectra and other observables. However, there is 
room for further improvement in comparison with experiment. 
In these calculations we used a strength of the 2$\pi$-exchange piece of the NNN interaction,
which is consistent with the NN interaction that we employed (i.e. from Ref.~\cite{N3LO}). 
This strength is somewhat uncertain (see e.g. Ref.~\cite{Nogga06}). Therefore, it will be
important to study the sensitivity of our results 
with respect to this strength.
Further on, it will be interesting to incorporate sub-leading NNN interactions and also
four-nucleon interactions, which are also order N$^3$LO \cite{Epelbaum06}.
Finally, it will be useful
to extend the basis spaces to  $N_{\rm max}=8$ ($8\hbar\Omega$) for $A>6$ 
to further improve convergence.

\section{Cluster overlap functions and S-factors of capture reactions}

In the {\it ab initio} NCSM calculations, we are able to obtain wave functions of
low-lying states of light nuclei in large model spaces. An interesting and important
question is, what is the cluster structure of these wave functions. That is we
want to understand, how much, e.g. an $^6$Li eigenstate looks like $^4$He plus deuteron,
an $^7$Be eigenstate looks like $^4$He plus $^3$He, an $^8$B eigenstate looks like
$^7$Be plus proton and so on. This information
is important for the description of low-energy nuclear reactions.
To gain insight, one introduces channel cluster form factors
(or overlap integrals, overlap functions). The formalism for calculating
the channel cluster form factors from the NCSM wave functions was developed in Ref.~\cite{cluster}.
Here we just briefly repeat a part of the formalism relevant to the simplest case 
when the lighter of the two clusters is a single-nucleon.   

We consider a composite system of $A$ nucleons, i.e. $^8$B, a nucleon projectile, here a proton,
and an $A-1$-nucleon target, i.e. $^7$Be.
Both nuclei are assumed to be described by eigenstates of the NCSM effective Hamiltonians
expanded in the HO basis with identical HO frequency and the same
(for the eigenstates of the same parity) or differing by one unit
of the HO excitation (for the eigenstates of opposite parity)
definitions of the model space. The target and the composite system is assumed to be described
by wave functions expanded in Slater determinant single-particle HO basis (that is obtained from
a calculation using a shell model code like Antoine).

Let us introduce a projectile-target wave function
\begin{eqnarray}\label{proj-targ_state_delta}
\langle\vec{\xi}_1 \ldots \vec{\xi}_{A-2} r^\prime \hat{r}
|\Phi_{(l\frac{1}{2})j;\alpha I_1}^{(A-1,1)J M};\delta_{r}\rangle
&=&\sum (j m I_1 M_1 | J M) (l m_l \textstyle{\frac{1}{2}} m_s | j m)
\frac{\delta(r-r^\prime)}{r r^\prime}
\nonumber \\
& & \times
Y_{l m_l}(\hat{r}) \chi_{m_s}
\langle \vec{\xi}_1 \ldots \vec{\xi}_{A-2} | A-1 \alpha I_1 M_1\rangle \; ,
\end{eqnarray}
where
$\langle \vec{\xi}_1 \ldots \vec{\xi}_{A-2} | A-1 \alpha I_1 M_1\rangle$
and
$\chi_{m_s}$ are the target and the nucleon wave function, respectively.
Here, $l$ is the channel relative orbital angular momentum, $\vec{\xi}$ 
are the target Jacobi coordinates defined in Eq.~(\ref{jacobiam11}) and
$\vec{r}=\left[\frac{1}{A-1}
      \left(\vec{r}_1+\vec{r}_2 + \ldots+ \vec{r}_{A-1}\right)-\vec{r}_{A}\right]$
describes the relative distance between the nucleon and the center of mass of the target.
The spin and isospin coordinates were omitted for simplicity.

The channel cluster form factor is then defined by
\begin{equation}\label{cluster_form_factor}
g^{A\lambda J}_{(l\frac{1}{2})j;A-1 \alpha I_1}(r)=
\langle A \lambda J |{\cal A}\Phi_{(l\frac{1}{2})j;\alpha I_1}^{(A-1,1)J};
\delta_{r}\rangle \; ,
\end{equation}
with ${\cal A}$ the antisymmetrizer and $|A\lambda J\rangle$ an eigenstate 
of the $A$-nucleon composite system (here $^8$B). It can be calculated from the
NCSM eigenstates obtained in the Slater-determinant basis from a
reduced matrix element of the creation operator. The derivation is as follows.
First, we use the relation (\ref{state_relation}) for both
the composite $A$-nucleon and the target $A-1$-nucleon
eigenstate. With the help of relations analogous to (\ref{HOtransform}):
\begin{eqnarray}\label{ho_tr}
&&\sum_{M m} (L M l m|Q q) \varphi_{N L M}(\vec{R}^{A-1}_{\rm CM}) 
\varphi_{n l m}(\vec{r}_A) = 
\nonumber \\
&&\sum_{n' l' m' N' L' M'} \langle n'l' N'L' Q|N L n l Q\rangle_{\frac{1}{A-1}}
(l' m' L' M'|Q q) \varphi_{n'l'm'}(\vec{\xi}_{A-1}) \varphi_{N'L'M'}(\vec{\xi}_0)
\; ,
\end{eqnarray}
we obtain
\begin{equation}\label{SD_Jacobi_overlap}
 _{\rm SD}\langle A\lambda J | {\cal A} 
\Phi_{(l\frac{1}{2})j;\alpha I_1}^{(A-1,1)J};nl\rangle_{\rm SD} \;
= \langle nl00l|00nll\rangle_{\frac{1}{A-1}} \;
\langle A\lambda J | {\cal A} 
\Phi_{(l\frac{1}{2})j;\alpha I_1}^{(A-1,1)J};nl\rangle
\; ,  
\end{equation}
with a general HO bracket due to the CM motion. The $nl$ in 
(\ref{SD_Jacobi_overlap}) refers to a replacement of $\delta_{r}$ by the HO 
$R_{nl}(r)$ radial wave function. 
Second, we relate the SD overlap to a linear combination of matrix elements 
of a creation operator 
between the target and the composite eigenstates
$_{\rm SD}\langle A\lambda J |a^\dagger_{nlj}| A-1 \alpha I_1 \rangle_{\rm SD}$. 
The subscript SD refers to the fact that these states were obtained
in the Slater determinant basis.
Such matrix elements are easily calculated by shell model codes.
The result is
\begin{eqnarray}\label{single-nucleon}
\langle A \lambda J|{\cal A} \Phi_{(l \textstyle{\frac{1}{2}},j);\alpha I_1}^{(A-1,1) J};
\delta_{r}\rangle
&=& \sum_n R_{nl}(r)
\frac{1}{\langle nl00l|00nll\rangle_{\frac{1}{A-1}}} \frac{1}{\hat{J}}
(-1)^{I_1-J-j}
\nonumber \\
&&\times
\; _{\rm SD}\langle A\lambda J||a^\dagger_{nlj}||A-1\alpha I_1\rangle_{\rm SD} \;
\; .
\end{eqnarray}
The eigenstates expanded in the Slater determinant basis contain CM components.
A general HO bracket, which value is simply given by
\begin{equation}\label{cm_ho_br}
\langle nl00l|00nll\rangle_{\frac{1}{A-1}} = (-1)^l
\left(\frac{A-1}{A}\right)^{\frac{2n+l}{2}}
\; ,
\end{equation}
then appears in Eq. (\ref{single-nucleon}) in order to remove these components.
The $R_{nl}(r)$ in Eq. (\ref{single-nucleon}) is the radial HO
wave function with the oscillator length parameter $b=\sqrt{\frac{\hbar}{\frac{A-1}{A}m\Omega}}$,
where $m$ is the nucleon mass.

A conventional spectroscopic factor is obtained by integrating the square of the cluster form
factor:
\begin{equation}\label{spec_fac}
S^{A\lambda J}_{(l\frac{1}{2})j;A-1 \alpha I_1}=
\int dr r^2
|g^{A\lambda J}_{(l\frac{1}{2})j;A-1 \alpha I_1}(r)|^2
\; .
\end{equation}

A generalization for projectiles (= the lighter of the two clusters) with 2, 3 or 4
nucleons is straightforward, although the expressions become more involved. In all cases,
the projectile is described by wave function expanded in Jacobi coordinate HO basis,
while the composite and the target eigenstates are expanded in the Slater determinant
HO basis. Full details are given in Ref.~\cite{cluster}.

\begin{figure}[htb]
\vspace{1.1cm}
\begin{center}
{\includegraphics[width=0.8\columnwidth]{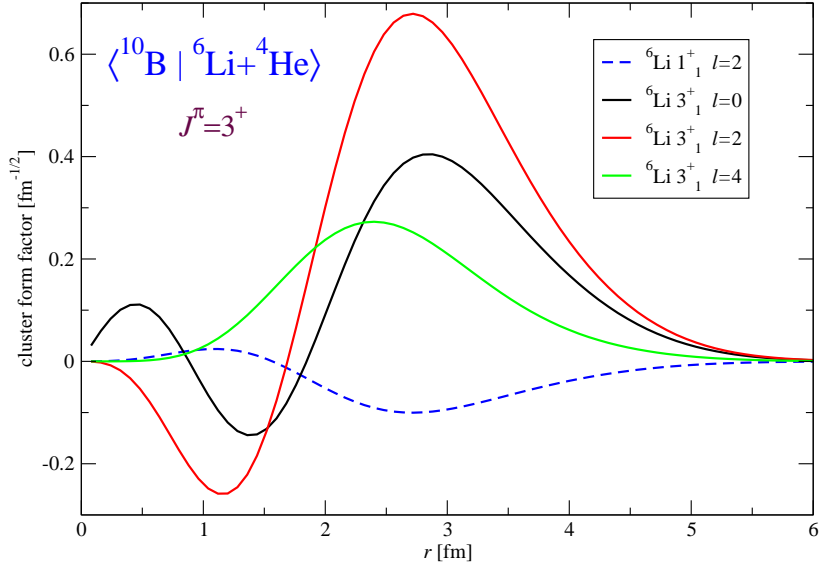}}
\caption{Overlap integral of $^{10}$B ground state with $^6$Li+$^4$He 
as a function of separation between the $^4$He and the$^6$Li. Results for 
the $^6$Li $1^+ 0$ ground state and the first excited $3^+ 0$ state are compared. 
The $\chi$EFT NN+NNN interaction and the $N_{max=6}$ model space for $^{10}$B and $^6$Li
were used.}
\label{B10_Li6_overlap}
\end{center}
\end{figure}

As an example, in Fig.~\ref{B10_Li6_overlap} we present the cluster overlap function
of $^{10}$B ground state with $^6$Li+$^4$He. Results are given for $^6$Li in the $1^+ 0$
ground state and in the $3^+ 0$ excited state. We can see that the  $^6$Li $1^+ 0$ ground-state
component is rather small. The $^{10}$B ground state is dominated by a superposition
of $S$-, $D$- and $G$-waves of relative motion of $^4$He and $^6$Li in the $3^+ 0$ state.

The overlap functions introduced in this subsection are relevant for description
of low-energy $\gamma$-capture reactions important for nuclear astrophysics. Next, we
investigate three reactions of this type.

\subsection{$^7$Be(p,$\gamma$)$^8$B}

The $^7$Be(p,$\gamma$)$^8$B capture reaction serves as an important
input for understanding the solar neutrino flux \cite{Adelberger}.
Recent experiments have determined the neutrino flux emitted from
$^8$B with a precision of ~9\% \cite{SNO}. On the other hand,
theoretical predictions have uncertainties of the order of 20\%
\cite{CTK03,BP04}. The theoretical neutrino flux depends on the
$^7$Be(p,$\gamma$)$^8$B S-factor. Many experimental and theoretical
investigations studied this reaction.

In this subsection, we discuss a calculation of the
$^7$Be(p,$\gamma$)$^8$B S-factor starting from {\it ab initio}
wave functions of $^8$B and $^7$Be.
It should be noted that the aim of {\it ab initio} approaches is to predict
correctly absolute cross sections (S-factors), not only relative cross sections.
The full details
of our $^7$Be(p,$\gamma$)$^8$B investigation were published in 
Refs.~\cite{Be7_p_NCSM_lett,Be7_p_NCSM}.

\begin{figure}[htb]
\vspace{1cm}
\begin{center}
  \includegraphics[width=0.8\columnwidth]
{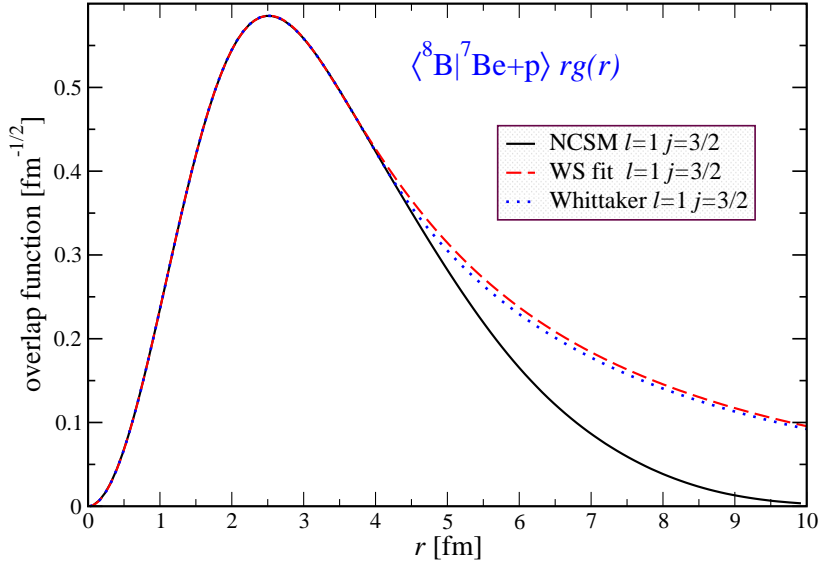}
\caption{\label{B8_Be7+p_overlap} 
Overlap function, $rg(r)$, for the ground state of $^8$B with the ground 
state of $^7$Be plus proton as a dependence on separation between the $^7$Be 
and the proton. The $p$-wave channel with $j=3/2$
is shown.
The full line represents the NCSM result obtained using the CD-Bonn 2000 NN
potential, the $10\hbar\Omega$ model space and the HO frequency 
of $\hbar\Omega=12$ MeV. The dashed lines
represent corrected overlaps obtained from a Woods-Saxon potential
whose parameters were fit to the NCSM overlaps up to ~4.0 fm under the 
constraint to reproduce the experimental separation energy. The dotted lines
represent overlap corrections by the direct Whittaker function matching. 
}
\end{center}
\end{figure}

Our calculations for both $^7$Be and $^8$B nuclei were performed
using the high-precision CD-Bonn 2000 NN potential \cite{cdb2k}
in model spaces up to $10\hbar\Omega$ ($N_{\rm max}=10$) for a wide range of HO frequencies.
From the obtained $^8$B and $^7$Be wave functions, we calculate the channel cluster
form factors (overlap functions, overlap integrals)
$g^{A\lambda J}_{(l\frac{1}{2})j;A-1 \alpha I_1}(r)$ as discussed in the previous subsection.
Here, $A=8$, $l$ is the channel
relative orbital angular momentum and
$\vec{r}=\left[\frac{1}{A-1}
      \left(\vec{r}_1+\vec{r}_2 + \ldots+ \vec{r}_{A-1}\right)-\vec{r}_{A}\right]$
describes the relative distance between the proton and the center of mass of
$^7$Be.
The two most important channels are the $p$-waves, $l=1$, with the proton
in the $j=3/2$ and $j=1/2$ states, $\vec{j}=\vec{l}+\vec{s}, s=1/2$. In these channels,
we obtain the spectroscopic factors of $0.96$ and $0.10$, respectively. The dominant
$j=3/2$ overlap integral is presented in 
Fig. \ref{B8_Be7+p_overlap} by the full line. The $10\hbar\Omega$ model space 
and the HO frequency of $\hbar\Omega=12$ MeV were used.
Despite the fact, that a very large basis was employed in the present calculation, 
it is apparent that the overlap function is nearly zero at about 10 fm. This is
a consequence of the HO basis asymptotic behavior. As already discussed, 
in the {\it ab initio} NCSM, the short-range correlations are taken into account
by means of the effective interaction. The medium-range correlations are then included
by using a large, multi-$\hbar\Omega$ HO basis. The long-range behavior is not treated
correctly, however. 
The proton capture on $^7$Be to the 
weakly bound ground state of $^8$B associated dominantly by the $E1$ radiation
is a peripheral process. In order to calculate the S-factor of this process we need
to go beyond the {\it ab initio} NCSM as done up to this point.
We expect, however, that the interior part of the overlap function
is realistic. It is then straightforward to find a quick fix and 
correct the asymptotic behavior of the overlap functions, which should be
proportional to the Whittaker function. 

One possibility we explored utilizes solutions of a Woods-Saxon (WS) potential.
In particular, we performed
a least-square fit of a WS potential solution to the interior of the
NCSM overlap in the range of $0-4$ fm. The WS potential parameters
were varied in the fit under the constraint that the experimental
separation energy of $^7$Be+p, $E_0=0.137$~MeV, was reproduced. In this way we obtain a perfect
fit to the interior of the overlap integral and a correct asymptotic behavior
at the same time. The result is shown in Fig. \ref{B8_Be7+p_overlap}
by the dashed line.

Another possibility is a direct matching of logarithmic derivatives of the NCSM overlap integral
and the Whittaker function:
$\frac{d}{dr}ln(rg_{lj}(r))=\frac{d}{dr}ln(C_{lj} W_{-\eta,l+1/2}(2k_0r))$,
where $\eta$ is the Sommerfeld parameter, $k_0=\sqrt{2\mu E_0}/\hbar$ with $\mu$ the reduced mass
and $E_0$ the separation energy.
Since asymptotic normalization constant (ANC) $C_{lj}$ cancels out, there 
is a unique solution at $r=R_m$.
For the discussed overlap presented in Fig.~\ref{B8_Be7+p_overlap}, 
we found $R_m=4.05$~fm.
The corrected overlap using the Whittaker function matching is shown 
in Fig.~\ref{B8_Be7+p_overlap}
by a dotted line. In general, we observe that the approach using the WS fit leads 
to deviations from the
original NCSM overlap starting at a smaller radius. In addition, the WS solution fit introduces
an intermediate range from about 4 fm to about 6 fm, where the corrected overlap deviates
from both the original NCSM overlap and the Whittaker function. Perhaps, this is a more realistic
approach compared to the direct Whittaker function matching. In any case, 
by considering the two alternative
procedures we are in a better position to estimate uncertainties in our S-factor results.

In the end, we re-scale the corrected overlap functions to preserve the original
NCSM spectroscopic factors (Table 2 of Ref.~\cite{Be7_p_NCSM_lett}).
In general, we observe a faster convergence of the spectroscopic
factors than that of the overlap functions. The corrected
overlap function should represent the infinite space result. By re-scaling
a corrected overlap function obtained at a finite $N_{\rm max}$, we approach
faster the infinite space result. At the same time, by re-scaling we preserve the
spectroscopic factor sum rules.

The S-factor for the reaction  $^7{\rm Be(p},\gamma)^8{\rm B}$
also depends on the continuum wave function,
$R_{lj}^{(c)}$. As we have not yet developed an
extension of the NCSM to describe continuum wave functions
(see, however, the discussion in Sect.~\ref{NCSM_RGM}),
we obtain  $R_{lj}^{(c)}$ for $s$ and $d$ waves from
a WS potential model.
Since the largest part of the integrand stays outside the
nuclear interior, one expects that the continuum wave functions are
well described in this way.
In order to have the same scattering wave function in all the calculations,
we chose a WS potential from Ref.~\cite{Esbensen} that was fitted to
reproduce the $p$-wave $1^+$ resonance in $^8$B.
It was argued \cite{Robertson}
that such a potential is also suitable for the description of $s$- and $d$-waves.
We note that the S-factor is very weakly dependent on the choice
of the scattering-state potential (using our fitted potential for the scattering state
instead changes the S-factor by less than 1.5 eV b at 1.6 MeV with no change at 0 MeV).

\begin{figure}[htb]
\vspace{1cm}
\begin{center}
  \includegraphics[width=0.8\columnwidth]{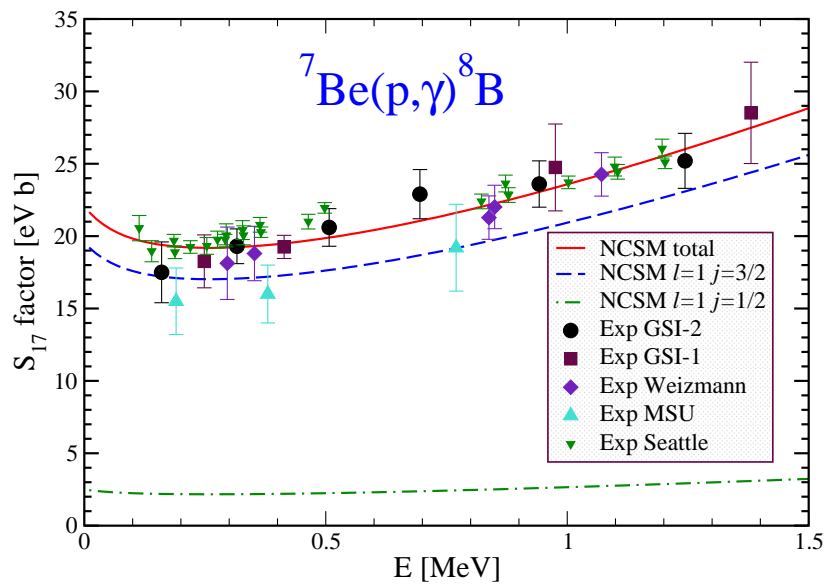}
  \caption{\label{S-factor_12_Nmax} 
The $^7$Be(p,$\gamma$)$^8$B S-factor obtained
using the NCSM overlap functions with corrected asymptotics 
as described in the text.
The dashed and dashed-dotted lines show the contribution due to the $l=1$, 
$j=3/2$ and $j=1/2$ partial waves, respectively.
Experimental values are from Refs. \protect\cite{Seattle,Be7pgamm_exp}.
}
\end{center}
\end{figure}

Our obtained S-factor is presented in Figs. \ref{S-factor_12_Nmax} 
where contribution
from the two partial waves are shown together with the total result. 
It is interesting
to note a good agreement of our calculated S-factor with the recent Seattle direct
measurement \cite{Seattle}.

In order to judge the convergence of our S-factor calculation, we performed
a detailed investigation of the model-space-size and the HO frequency dependencies.
We used the HO frequencies in the range from
$\hbar\Omega=11$ MeV  to $\hbar\Omega=15$ MeV and the model spaces from 
$6\hbar\Omega$ to $10\hbar\Omega$.
By analysing these results, we arrived at the S-factor value of
$S_{17}(10\;{\rm keV})=22.1\pm 1.0$ eV b.

\subsection{$^3$He($\alpha$,$\gamma$)$^7$Be}

\begin{figure}[htb]
\vspace{1cm}
\begin{center}
  \includegraphics[width=0.8\columnwidth]{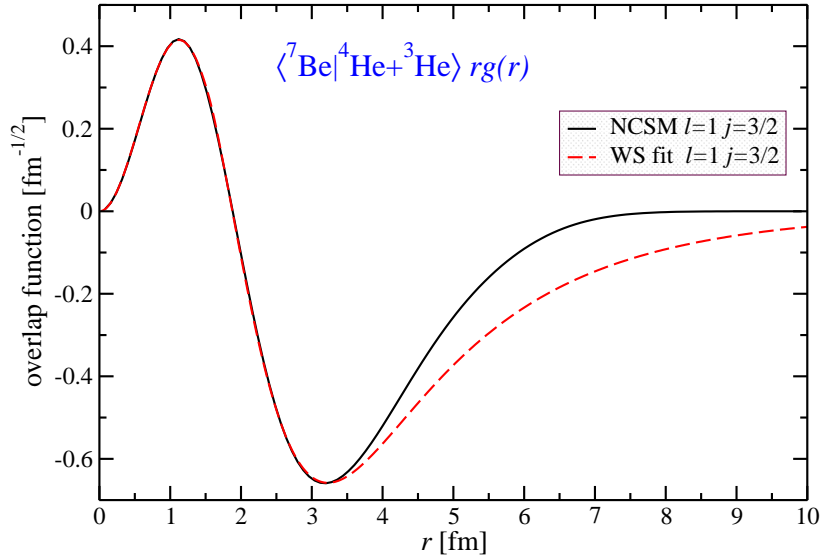}
  \caption{\label{overlap_34} 
The overlap function, $rg(r)$, for the first excited state of $^7$Be with the ground 
state of $^3$He plus $\alpha$ as a dependence on separation between the $^3$He 
and the $\alpha$ particle. The $p$-wave channel overlap function with $j=3/2$
is shown.
The full line represents the NCSM result obtained using the CD-Bonn 2000 NN 
potential and the $10\hbar\Omega$
model space for $^7$Be ($12\hbar\Omega$ for $^{3,4}$He) 
with the HO frequency of $\hbar\Omega=13$ MeV. The dashed line
represents a corrected overlap obtained with a Woods-Saxon potential
whose parameters were fit to the NCSM overlap up to ~3.4 fm under the 
constraint to reproduce the experimental separation energy. 
}
\end{center}
\end{figure}

The $^3$He($\alpha$,$\gamma$)$^7$Be capture reaction cross section was
identified the most important uncertainty in the solar model predictions 
of the neutrino fluxes in the p-p chain \cite{BP04}. 
We investigated the bound states
of $^7$Be, $^3$He and $^4$He within the {\it ab initio} NCSM and calculated the overlap
functions of $^7$Be bound states with the ground 
states of $^3$He plus $^4$He as a function of separation between the $^3$He 
and the $\alpha$ particle.
The obtained $p$-wave overlap functions of the $^7$Be
$3/2^-$ ground state 
excited state are presented in Fig.~\ref{overlap_34} by the full line.
The dashed lines show the corrected overlap function obtained by the least-square fits
of the WS parameters done in the same way as in the $^8$B$\leftrightarrow ^7$Be+p case.
The corresponding NCSM spectroscopic factors obtained using the CD-Bonn 2000 in the
$10\hbar\Omega$ model space for $^7$Be ($12\hbar\Omega$ for $^{3,4}$He) 
and HO frequency of $\hbar\Omega=13$ MeV are 0.93 and 0.91
for the ground state and the first excited state of $^7$Be, respectively. We note 
that contrary to
the $^8$B$\leftrightarrow ^7$Be+p case, the $^7$Be$\leftrightarrow ^3$He+$\alpha$ 
$p$-wave overlap functions have a node.

\begin{figure}[htb]
\vspace{1cm}
\begin{center}
  \includegraphics[width=0.8\columnwidth]{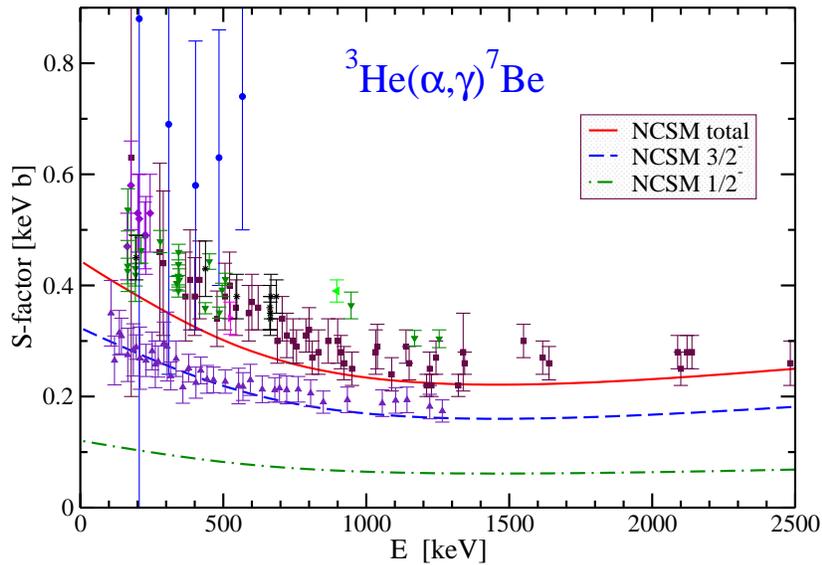}
  \caption{\label{S-factor_34_Nmax}
The full line shows the $^3$He($\alpha$,$\gamma$)$^7$Be S-factor obtained
using the NCSM overlap functions with corrected asymptotics. The dashed lines
show the $^7$Be ground- and the first excited state contributions.
The calculation was done using the CD-Bonn 2000 NN 
potential and the $10\hbar\Omega$ model space for $^7$Be 
($12\hbar\Omega$ for $^{3,4}$He) 
with the HO frequency of $\hbar\Omega=13$ MeV.  
}
\end{center}
\end{figure}

Using the corrected overlap functions and a $^3$He+$\alpha$ 
scattering state obtained using the potential
model of Ref.~\cite{Kim} we calculated the $^3$He($\alpha$,$\gamma$)$^7$Be S-factor.
Our $10\hbar\Omega$ result is presented in the left panel of Fig.~\ref{S-factor_34_Nmax}.
We show the total S-factor as well as the contributions from the capture to the ground
state and the first excited state of $^7$Be.
By investigating the model space dependence for $8\hbar\Omega$ and $10\hbar\Omega$
spaces 
we estimate the $^3$He($\alpha$,$\gamma$)$^7$Be S-factor at zero energy 
to be higher than 0.44 keV b, the value that we obtained in the discussed
case shown in Fig.~\ref{S-factor_34_Nmax}.
Our results are similar to those obtained by K. Nollett \cite{Nollett}
using the variational Monte Carlo wave functions for the bound states and 
potential model wave functions for the scattering state.

\subsection{$^3$H($\alpha$,$\gamma$)$^7$Li}

\begin{figure}[htb]
\vspace{1cm}
\begin{center}
  \includegraphics[width=0.8\columnwidth]{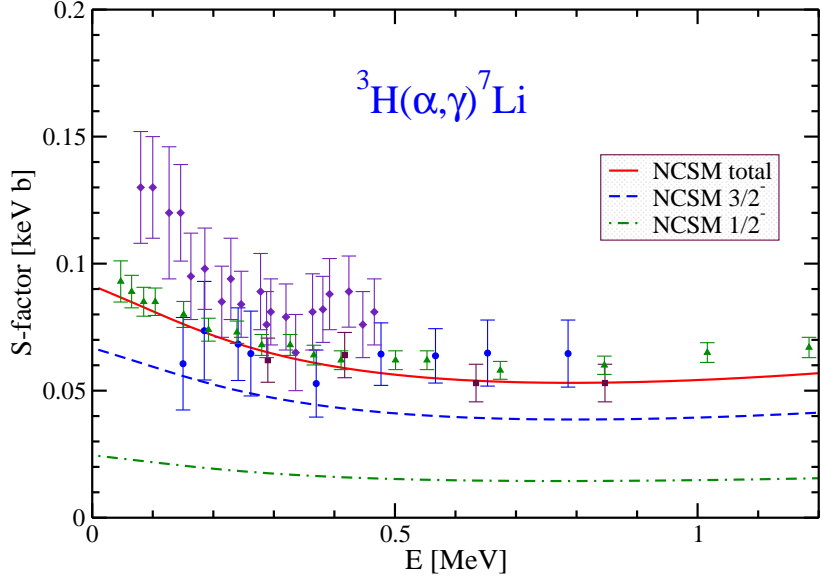}
  \caption{\label{S-factor_Li7_Nmax}
The full line shows the $^3$H($\alpha$,$\gamma$)$^7$Li S-factor obtained
using the NCSM overlap functions with corrected asymptotics. The dashed lines
show the $^7$Li ground- and the first excited state contributions.
The calculation was done using the CD-Bonn 2000 NN 
potential and the $10\hbar\Omega$ model space for $^7$Li 
($12\hbar\Omega$ for $^3$H and $^{4}$He) 
with the HO frequency of $\hbar\Omega=13$ MeV.  
}
\end{center}
\end{figure}

An important check on the consistency of the $^3$He($\alpha$,$\gamma$)$^7$Be S-factor
calculation is the investigation of the mirror reaction $^3$H($\alpha$,$\gamma$)$^7$Li,
for which more accurate data exist \cite{Brune}. Our results obtained using the CD-Bonn 2000
NN potential are shown in Fig.~\ref{S-factor_Li7_Nmax}. It is apparent that
our $^3$H($\alpha$,$\gamma$)$^7$Li results are consistent with our 
$^3$He($\alpha$,$\gamma$)$^7$Be calculation. We are on the lower side of the data
and we find an increase of the S-factor as we increase the size of our basis.

More details on the {\it ab initio} NCSM investigation of the $^3$He($\alpha$,$\gamma$)$^7$Be
and $^3$H($\alpha$,$\gamma$)$^7$Li S-factors are given in Ref.~\cite{S_fact_NPA}.

\section{Towards the {\it ab initio} NCSM with continuum}\label{NCSM_RGM}

In the previous section, we highlighted shortcomings of the {\it ab initio} NCSM, 
its incorrect description of long-range correlations and its lack of coupling to continuum.
If we want to build upon the {\it ab initio} NCSM to microscopically describe loosely
bound systems as well as nuclear reactions, the approach must be augmented by explicitly
including cluster states such as, e.g. those given in Eq.~(\ref{proj-targ_state_delta}), 
and solve for their relative motion while imposing the proper boundary conditions. This can be
done by extending the {\it ab initio} NCSM HO basis through the addition of 
the cluster states. This would result
in an over-complete basis with the cluster relative motion wave functions as amplitudes
that need to be determined. The first step in this direction is to consider the cluster
basis alone. This approach is very much in the
spirit of the resonating group method (RGM) \cite{RGM}, 
a technique that considers clusters with fixed
internal degrees of freedom, treats the Pauli principle exactly and solves the many-body problem
by determining the relative motion between the various clusters. In our approach, we use
the {\it ab initio} NCSM wave functions for the clusters involved and the {\it ab initio} NCSM 
effective interactions derived from realistic NN (and eventually also from NNN) potentials.

The general outline of the formalism is as follows. The many-body wave function is
approximated by a superposition of binary cluster channel wave functions
\begin{equation}\label{RGM_wave}
\Psi^{(A)}=\sum_{\nu}\hat{\mathcal A}
\left[\psi_{1\nu}^{(A-a)}\psi_{2\nu}^{(a)}\varphi_{\nu}(\vec{r}_{A-a,a})\right]
=\sum_{\nu}\int d\vec{r}\,\varphi_{\nu}(\vec{r}\,)\,\hat{\mathcal A}\,\Phi_{\nu\vec{r}}^{(A-a,a)}\,,
\end{equation}
with
\begin{equation}\label{RGM_basis}
\Phi_{\nu\vec{r}}^{(A-a,a)}=\psi_{1\nu}^{(A-a)}\psi_{2\nu}^{(a)}\delta(\vec{r}\,-\vec{r}_{A-a,a})\,.
\end{equation}
Here, $\hat{\mathcal A}$ is the antisymmetrizer accounting for the exchanges of nucleons 
between the two clusters (which are already antisymmetric with respect to exchanges
of internal nucleons). 
The relative-motion wave functions $\varphi_{\nu}$ depend on
the relative-distance between the center of masses of the two clusters in channel $\nu$. 
They can be determined by solving the many-body Schr\"{o}dinger
equation in the Hilbert space spanned by the basis functions (\ref{RGM_basis}):
\begin{equation}
H\Psi^{(A)}=E\Psi^{(A)}\longrightarrow\, \sum_{\nu}\int d\vec{r}\,
\left[{\mathcal H}^{(A-a,a)}_{\mu\nu}(\vec{r}\,^\prime,\vec{r}\,)
-E{\mathcal N}^{(A-a,a)}_{\mu\nu}(\vec{r}\,^\prime,\vec{r}\,)\right]\varphi_{\nu}(\vec{r}\,)\,,
\end{equation}
where the Hamiltonian and norm kernels are defined as
\begin{eqnarray}
{\mathcal H}^{(A-a,a)}_{\mu\nu}(\vec{r}\,^\prime,\vec{r}\,)
&=&\left\langle\Phi_{\mu\vec{r}\,^\prime}^{(A-a,a)}\left|\hat{\mathcal A}\,H\,
\hat{\mathcal A}\right|\Phi_{\nu\vec{r}}^{(A-a,a)}\right\rangle\,,\\
{\mathcal N}^{(A-a,a)}_{\mu\nu}(\vec{r}\,^\prime,\vec{r}\,)&=&
\left\langle\Phi_{\mu\vec{r}\,^\prime}^{(A-a,a)}\left|\hat{\mathcal A}^2\right|
\Phi_{\nu\vec{r}}^{(A-a,a)}\right\rangle\,.
\end{eqnarray}

The most challenging task is to evaluate the Hamiltonian kernel and the norm kernel. 
We now briefly outline,
how this is done when {\it ab initio} NCSM wave functions are used for the
binary cluster states. 
From now on, let us consider the cluster states with a single-nucleon projectile
($a=1$ in Eq.~\ref{RGM_wave}). A generalization
is straightforward. Using an alternative coupling scheme compared to 
Eq.~(\ref{proj-targ_state_delta}), we introduce
\begin{eqnarray}\label{proj-targ_state_delta_rgm}
&&\langle\vec{\xi}_1 \ldots \vec{\xi}_{A-2} \xi_{A-1}^\prime \hat{\xi}_{A-1}
|\Phi_{(\alpha I_1T_1,\frac{1}{2}\frac{1}{2});sl}^{(A-1,1)J M T M_T};\delta_{\xi_{A-1}}\rangle
\nonumber \\
&=&\sum (I_1 M_1 \textstyle{\frac{1}{2}} m_s| s m) (s m l m_l | J M)
(T_1 M_{T_1} \textstyle{\frac{1}{2}} m_t| T M_T)
\frac{\delta(\xi_{A-1}-\xi_{A-1}^\prime)}{\xi_{A-1} \xi_{A-1}^\prime}
\nonumber \\
&\times&
Y_{l m_l}(\hat{\xi}_{A-1}) \chi_{m_s} \chi_{m_t}
\langle \vec{\xi}_1 \ldots \vec{\xi}_{A-2} | A-1 \alpha I_1 M_1 T_1 M_{T_1} \rangle \; ,
\end{eqnarray}
with the spin and isospin coordinates omitted to simplify the notation. The Jacobi coordinates
were defined in Eq.~(\ref{jacobiam11}). 
Using the latter cluster basis and the following definition of the antisymmetrizer
$\hat{\mathcal A}=1/\sqrt{A}(1-\sum_{j=1}^{A-1} P_{j,A})$ with $P_{j,A}$ 
the transposition operator of nucleons $j$ and $A$, the norm kernel can be expressed as 
\begin{eqnarray}\label{norm_kernel}
{\mathcal N}^{(A-1,1)}_{\mu\nu}(r^\prime,r)&=&\delta_{\mu\nu}\frac{\delta(r^\prime-r)}{r^\prime r}
-(A-1)\sum_{n^\prime n} R_{n^\prime l^\prime}(r^\prime)
\nonumber \\
&\times&
\langle \Phi_{(\alpha' I'_1 T'_1,\textstyle{\frac{1}{2}}
\textstyle{\frac{1}{2}})s' l'}^{(A-1,1) JT};n'l'|P_{A,A-1}|
\Phi_{(\alpha I_1 T_1,\textstyle{\frac{1}{2}}
\textstyle{\frac{1}{2}})s l}^{(A-1,1) JT};nl\rangle R_{nl}(r) \; ,
\end{eqnarray}
with $\mu\equiv (\alpha' I'_1 T'_1,\textstyle{\frac{1}{2}} \textstyle{\frac{1}{2}})s'$,
$\nu\equiv (\alpha I_1 T_1,\textstyle{\frac{1}{2}} \textstyle{\frac{1}{2}})s$ and $P_{A,A-1}$
the transposition operator of nucleons $A$ and $A-1$. The coordinates $r$ are related to
$\xi_{A-1}$ by $r=\sqrt{\frac{A}{A-1}}\xi_{A-1}$ and the HO length parameter 
of the radial HO wave functions is $b=\sqrt{\frac{\hbar}{\frac{A-1}{A}m\Omega}}$.
The matrix element of the transposition operator $P_{A,A-1}$ can be directly evaluated
using the {\it ab initio} NCSM wave functions expanded in Jacobi coordinate HO basis 
following a procedure analogous to the derivation of Eq.~(\ref{t13t23}). 
However, a crucial feature of the {\it ab initio} NCSM approach is
that the matrix elements that enter the norm kernel and the Hamiltonian kernel can be equivalently
evaluated using the {\it ab initio} NCSM wave functions expanded in the Slater 
determinant HO basis. This is achieved in two stages. 
First, we calculate the SD matrix element as
\begin{eqnarray}\label{P_AA-1_SD}
&& _{\rm SD}\langle \Phi_{(\alpha' I'_1 T'_1,\textstyle{\frac{1}{2}}
\textstyle{\frac{1}{2}})s' l'}^{(A-1,1) JT};n'l'|P_{A,A-1}|
\Phi_{(\alpha I_1 T_1,\textstyle{\frac{1}{2}}
\textstyle{\frac{1}{2}})s l}^{(A-1,1) JT};nl\rangle_{\rm SD}  = 
\nonumber \\
&& \frac{1}{A-1} \sum_{jj'K\tau} 
\left\{ \begin{array}{ccc} I_1 & \textstyle{\frac{1}{2}} & s \\
  l & J & j    
\end{array}\right\}
\left\{ \begin{array}{ccc} I'_1 & \textstyle{\frac{1}{2}} & s' \\
  l' & J & j'    
\end{array}\right\}
\left\{ \begin{array}{ccc} I_1 & K & I'_1 \\
  j' & J & j    
\end{array}\right\}
\left\{ \begin{array}{ccc} T_1 & \tau & T'_1 \\
 \textstyle{\frac{1}{2}} & T & \textstyle{\frac{1}{2}}
\end{array}\right\}
\nonumber \\
&\times& \hat{s}\hat{s}'\hat{j}\hat{j}'\hat{K}\hat{\tau} (-1)^{I'_1+j'+J} 
(-1)^{T_1+\textstyle{\frac{1}{2}}+T} 
\nonumber \\
&\times&
_{\rm SD}\langle A-1 \alpha' I'_1 T'_1 ||| (a^\dagger_{nlj\textstyle{\frac{1}{2}}}
\tilde{a}_{n'l'j'\textstyle{\frac{1}{2}}})^{(K\tau)} ||| A-1 \alpha I_1 T_1 \rangle_{\rm SD}
\; .
\end{eqnarray}
Second, it is possible to show that
the matrix element in the SD basis is related to the one in the Jacobi 
coordinate basis:
\begin{eqnarray}\label{P_A-aA_SD_Jacobi}
&& _{\rm SD}\langle \Phi_{(\alpha' I'_1 T'_1,\textstyle{\frac{1}{2}}
\textstyle{\frac{1}{2}})s' l'}^{(A-1,1) JT};n'l'|P_{A,A-1}|
\Phi_{(\alpha I_1 T_1,\textstyle{\frac{1}{2}}
\textstyle{\frac{1}{2}})s l}^{(A-1,1) JT};nl\rangle_{\rm SD}  = 
\nonumber \\
&& \sum_{n_rl_rn'_rl'_rJ_r} \langle \Phi_{(\alpha' I'_1 T'_1,\textstyle{\frac{1}{2}}
\textstyle{\frac{1}{2}})
s' l'_r}^{(A-1,1) J_r T};n'_r l'_r | P_{A,A-1} |
\Phi_{(\alpha I_1 T_1,\textstyle{\frac{1}{2}}\textstyle{\frac{1}{2}})
s l_r}^{(A-1,1) J_r T};n_r l_r \rangle
\nonumber \\
&\times& \sum_{NL} \hat{l}\hat{l}'\hat{J}_r^2 (-1)^{s+l_r-s-l'_r}
\left\{ \begin{array}{ccc} s & l_r & J_r \\
  L & J & l    
\end{array}\right\}
\left\{ \begin{array}{ccc} s' & l'_r & J_r \\
  L & J & l'    
\end{array}\right\}
\nonumber \\
&\times&
\langle n_r l_r N L l | 0 0 n l l \rangle_{\frac{1}{A-1}}
\langle n'_r l'_r N L l' | 0 0 n' l' l' \rangle_{\frac{1}{A-1}}   \; .
\end{eqnarray}
This relation then defines a matrix that one inverts to get the Jacobi-coordinate
matrix element. This is analogous to what was done to obtain the translationally invariant
density in Ref.~\cite{tr_dens}.
The Hamiltonian kernel can be evaluated in a similar yet more involved way.
It consists of a kinetic term, a NN potential direct term associated 
with the operator $V_{A,A-1}(1-P_{A,A-1})$
and a NN potential exchange term associated with the operator $V_{A,A-2}P_{A,A-1}$ 
(plus terms arising from the NNN interaction).
The ability to employ wave functions expanded in the SD basis
opens the possibility to apply this formalism for nuclei with $A>5$.

\begin{figure}
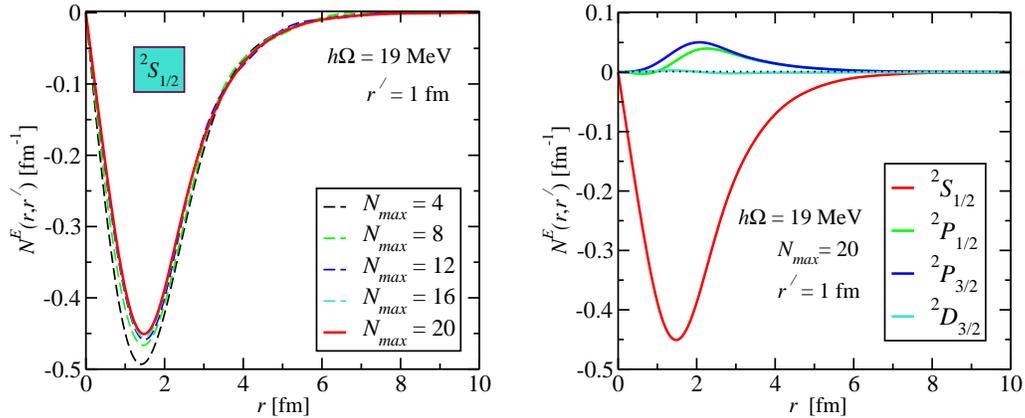

\vspace{1cm}
\begin{minipage}{7cm}
  \includegraphics[height=5.5cm]{He4-n-1p1.eps}
\end{minipage}
\hfill
\begin{minipage}{7cm}
  \includegraphics[height=5.5cm]{He4-n_2.eps}
\end{minipage}
  \caption{\label{Norm_kernel_n_4He} The exchange part of the norm kernel
of the n+$^4$He system. Left, the convergence with the size of the basis
of the $^4$He wave function for the $^{2}S_{1/2}$ channel. Right, results
for channels are compared. The chiral EFT NN potential was used.
}
\end{figure}

In Fig.~\ref{Norm_kernel_n_4He}, we show the exchange part of the norm kernel 
for the n+$^4$He system,
in particular the second term of Eq.~(\ref{norm_kernel}) multiplied by $r r^\prime$. 
It is apparent that we are able to reach convergence
for the kernel. Furthemore, the $^{2}S_{1/2}$ channel shows, as it should
the effect of the Pauli principle.
Indeed, the $^4$He wave function is dominated by the four nucleon $s$-shell configuration.
The Pauli principle prevents adding the fifth nucleon to the same shell.

\begin{figure}[htb]
\vspace{1.5cm}
\begin{center}
  \includegraphics[width=0.7\columnwidth]{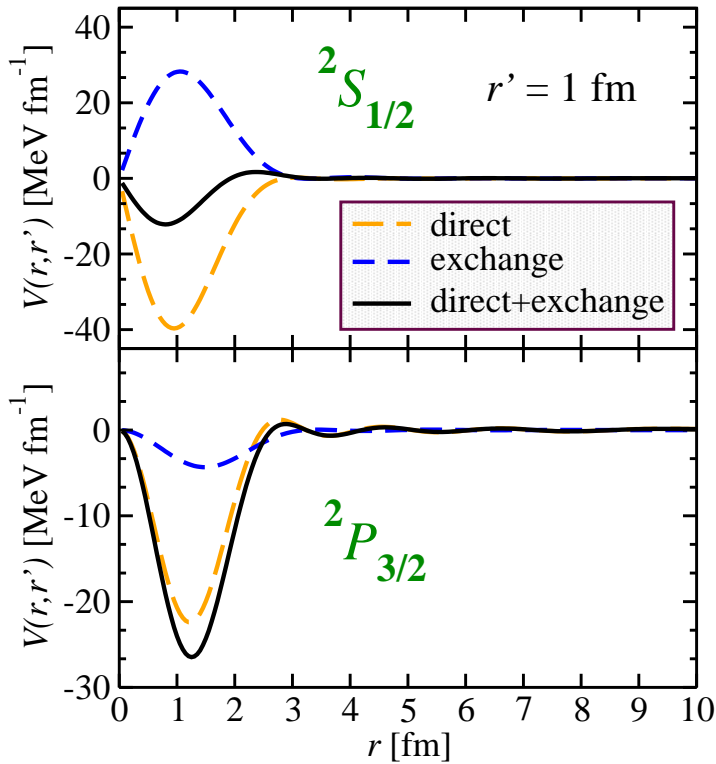}
  \caption{\label{potential_kernel_n_4He}
The NN potential direct and exchange terms of the Hamiltonian kernel of the
n+$^4$He system. The $^{2}S_{1/2}$ and $^{2}P_{3/2}$ channels are compared.
The low-momentum NN potential of Ref.~\cite{BoKu03} was used
in the $14\hbar\Omega$ model space with the HO frequency of $\hbar\Omega=18$~MeV.
}
\end{center}
\end{figure}

\begin{figure}[htb]
\vspace{1cm}
\begin{center}
  \includegraphics[width=0.7\columnwidth,height=8cm]{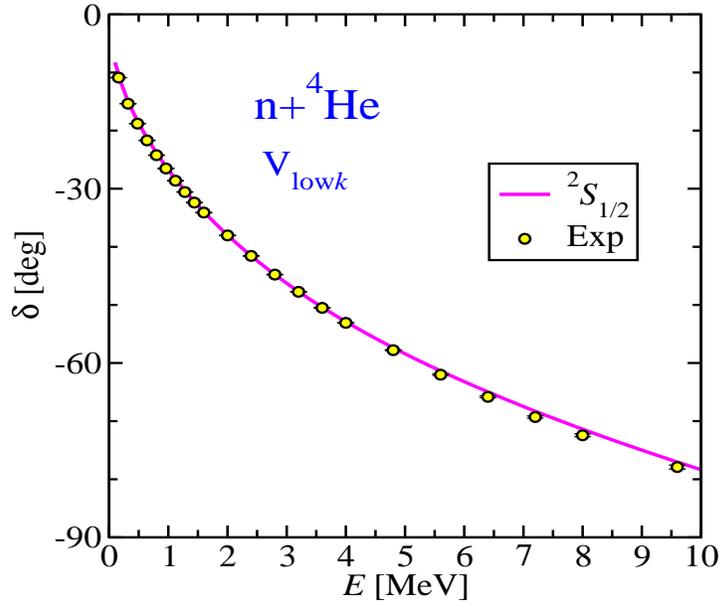}
  \caption{\label{n_4He_phase_shift}
The calculated $^{2}S_{1/2}$ phase shift for the n+$^4$He system compared with
experimental data. The low-momentum NN potential of Ref.~\cite{BoKu03} was used
in the $16\hbar\Omega$ model space with the HO frequency of $\hbar\Omega=18$~MeV.
}
\end{center}
\end{figure}

In Fig.~\ref{potential_kernel_n_4He}, we show the direct and the exchange contributions
of the NN potential to 
the Hamiltonian kernel as well as their sum for the n+$^4$He system. Again,
the Pauli principle is manifest in the $^{2}S_{1/2}$ channel. We were able to obtain the
presented results using wave functions expanded both in the Jacobi-coordinate and the SD basis.
The two independent calculations gave identical results as expected. 
A converged calculation of the $^{2}S_{1/2}$ phase shift together with experimental data
is presented in Fig.~\ref{n_4He_phase_shift}.
Full details regarding this approach are given in Ref.~\cite{NCSM_RGM}.

\section{Conclusions}

The {\it ab initio} NCSM evolved into a powerful many-body technique. Presently,
it is the only method capable to use interactions derived within the chiral EFT
for systems of more than four nucleons, in particular for mid-$p$-shell nuclei.
Among its successes is the demonstration of importance of the NNN interaction
for nuclear structure. Applications to nuclear reactions with a proper treatment
of long-range properties are under development.
Extension to heavier nuclei is achieved through the importance-truncated NCSM \cite{Imp_tr_NCSM}.
Within this approach, {\it ab initio} calculations for nuclei as heavy as $^{40}$Ca 
become possible.

\acknowledgments
I would like to thank all the collaborators that contributed to the cited 
papers and, in particular, Sofia Quaglioni for useful discussions and input
for section 5. 
This work performed under the auspices of the U.S. Department of Energy 
by Lawrence Livermore National Laboratory under Contract DE-AC52-07NA27344.
Support from the LDRD contract No.~04--ERD--058 and from
U.S. DOE/SC/NP (Work Proposal Number SCW0498) is acknowledged.
This work was also supported in part by the Department of Energy under
Grant DE-FC02-07ER41457.

\end{document}